%% file: main.tex
\documentclass[twocolumn,secnumarabic,amssymb,nobibnotes,aps,prl,superscriptaddress]{revtex4-2}

\usepackage{amsmath}
\usepackage{amsfonts}
\usepackage{amssymb}
\usepackage{graphicx}
\usepackage{hyperref}
\usepackage{longtable}
\usepackage{dcolumn}
\usepackage{bm}
\usepackage{epstopdf}
\usepackage{datetime}
\longdate
\usepackage{color}

\begin{document}

\title{A Twist On Active Membranes: Odd Mechanics, Spontaneous Flows and Shape Instabilities}

\author{Sami C. Al-Izzi}
\email{samiali@uio.no}
\affiliation{Department of Mathematics, Faculty of Mathematics and Natural Sciences, University of Oslo, 0315 Oslo, Norway}  

\author{Gareth P. Alexander}
\email{g.p.alexander@warwick.ac.uk}
\affiliation{Department of Physics and Centre for Complexity Science, University of Warwick, Coventry, CV4 7AL, United Kingdom}

\date{\today}

\begin{abstract}
Living systems are chiral on multiple scales, from constituent biopolymers to large scale morphology, and their active mechanics is both driven by chiral components and serves to generate chiral morphologies. 
We describe the mechanics of active fluid membranes in coordinate-free form, with focus on chiral contributions to the stress. These generate geometric `odd elastic' forces in response to mean curvature gradients but directed perpendicularly. As a result, they induce tangential membrane flows that circulate around maxima and minima of membrane curvature. When the normal viscous force amplifies perturbations the membrane shape can become linearly unstable giving rise to shape instabilities controlled by an active Scriven-Love number. We describe examples for spheroids, membranes tubes and helicoids, discussing the relevance and predictions such examples make for a variety of biological systems from the sub-cellular to tissue level.
\end{abstract}

\maketitle

The mechanics of active materials is distinct from that of passive systems: active stresses and forces are used to enable motility, sustain flows and coordinate a multitude of cellular functions essential for all living tissues. Theories of biological tissues as active gels and active liquid crystals have been successful in accounting for the hydrodynamics of the cytoskeleton, actin cortex and a variety of tissues, cell populations and biofilms~\cite{prost2015,julicher2018,saw2018}. 
In many cases these tissues are thin and deformable, and may be modelled as a two-dimensional surface whose active mechanics influences its shape~\cite{bailles2019,gehrels2023,khoromskaia2023,morris2019,salbreux2017,salbreux2022,mietke2019,al-izzi2021,al-izzi2023}.

Recently, the unique characteristics of chiral active mechanics has come to the fore~\cite{scheibner2020,kole2021,tan2022,furthauer2013,kole2023}. A striking example is provided by developing starfish embryos, which self-assemble into rotating crystals displaying odd mechanics~\cite{tan2022}. 
Odd elastic materials are characterised by non-reciprocal interactions that facilitate oscillations, work cycles, and sustained locomotion~\cite{fruchart2023,chen2021,ishimoto2022,fossati2022,scheibner2020,brandenbourger2021,kole2023}. 
More generally we note that chiral structures are commonplace in biology, with well-known examples including rotary motors~\cite{mandadapu2015}, DNA~\cite{worcel1981} and iridescence~\cite{sharma2009}, while left-right symmetry breaking is a basic feature of morphogenesis and development~\cite{naganathan2014,naganathan2016,taniguchi2011,inaki2018}. Of particular note with regards to left-right symmetry breaking are examples in the model organisms \textit{C. elegans}, where chiral flows in actomyosin occur in the zygote before the principal division~\cite{naganathan2014}, and \textit{Drosophila melanogaster}, where chiral stresses in the hind-gut cause it to twist into its characteristic ``question mark'' shape~\cite{taniguchi2011}.

Here we describe the active mechanics of fluid membranes in geometric, coordinate-free form, with focus on chiral aspects of the activity. The chiral activity contributes a tangential membrane force in response to gradients of mean curvature but acting at right angles to them, in a geometric counterpart of odd elasticity~\cite{scheibner2020,fruchart2023}. This is an odd response to bending elasticity, where the force acts along a direction that is fluid and hence drives membrane flows. We describe general linear deformations of spherical membranes, where the `odd' flows are induced directly by the bending mode, and also the steady odd flows generated by the geometry of spheroids with arbitrary aspect ratio. 

In addition to the induced odd membrane flow, the viscous force accompanying it has a normal component -- the Scriven-Love term -- that can amplify the curvature gradients generating the flow, resulting in a linear shape instability. We analyse this for the case of membrane tubes, where the instability leads to chiral symmetry breaking of the tube shape in the growth of a helical deformation with handedness set by the chiral activity. Finally, we also describe the odd mechanics of helicoids, for which the `odd' flows drive a shape instability qualitatively similar to that of soap films~\cite{boudaoud1999,machon2016PRL}, and of `membrane ramps' used to describe the geometry of the endoplasmic reticulum and plant photosynthetic membrane~\cite{teresaki2013,guven2014,dasilva2021,bussi2019}.

\begin{figure*}[t!]
\centering
\includegraphics[width=\textwidth]{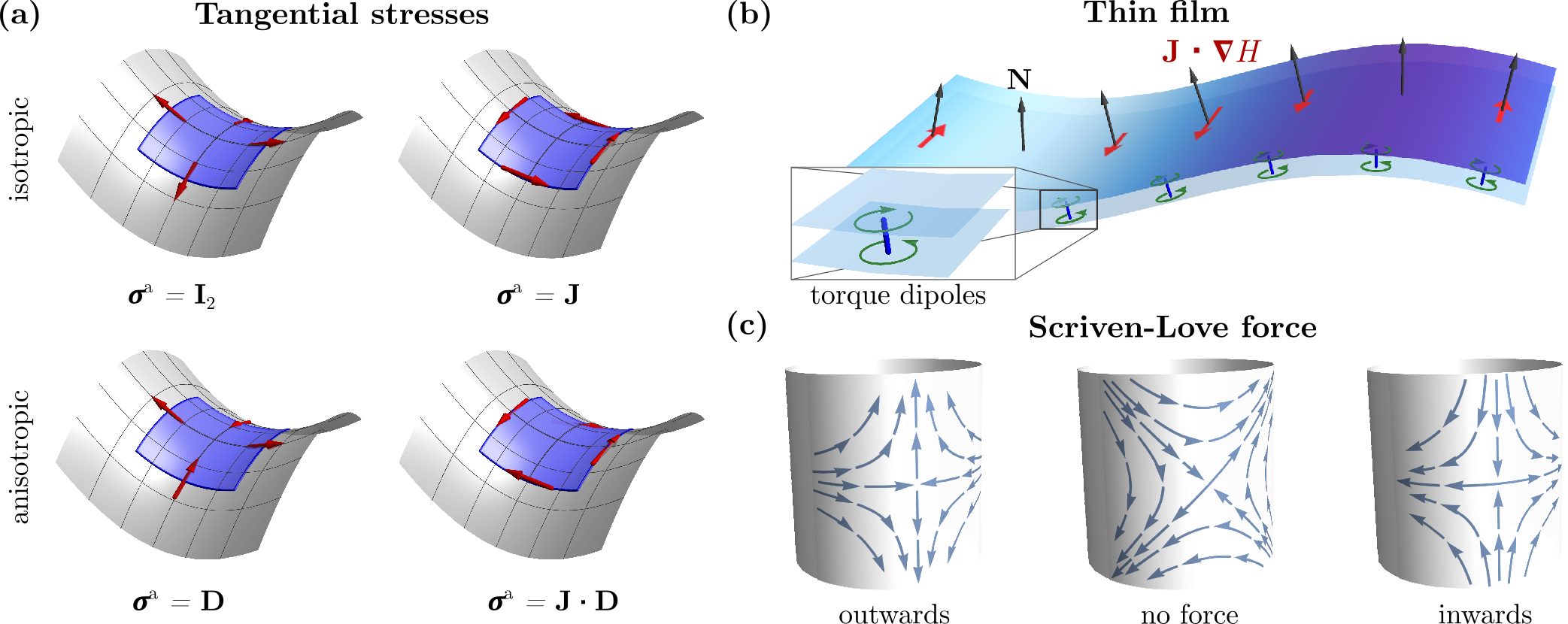}
\caption{\textbf{Schematic illustration of membrane mechanics.} \textbf{(a)} The basic tangential stresses ${\bf I}_2$, ${\bf J}$, ${\bf D}$, and ${\bf J}\cdot{\bf D}$. The arrows (red) indicate the forces acting on the highlighted patch (blue) from each part of its boundary. The mesh is aligned with the principal curvature directions of the surface. \textbf{(b)} Schematic of a thin chiral active film illustrating how torque dipoles (inset) can give rise to the force $\sim{\bf J}\cdot\nabla H$ (red arrows), that is forces perpendicular to gradients in mean curvature. The colour represents the variation in mean curvature. \textbf{(c)} The Scriven-Love force arising from the interaction of viscous stresses and membrane curvature. An extensional flow aligned with the principal curvature directions generates a normal force; at $45^{\circ}$ to the principal curvature axes there is no force.}
\label{fig:stresses}
\end{figure*}

\textbf{Membrane geometry and forces:} We represent the membrane by a surface smoothly embedded in $\mathbb{R}^3$ and describe its geometry using the surface gradient of the normal vector ${\bf N}$, known as the shape operator ${\bf S} = - \nabla {\bf N}$~\cite{doCarmo,Kobayashi,GrayAbbenaSalamon,KarcherNotes}. This is a linear transformation on the tangent space, whose eigenvalues $k_1, k_2$ are the principal curvatures and eigenvectors the principal curvature directions. The mean curvature $H = \frac{1}{2} \,\textrm{tr}\,{\bf S} = \frac{1}{2}(k_1+k_2)$ and Gaussian curvature $K_{\textrm{G}} = \det {\bf S} = k_1 k_2$ are the two scalar curvature invariants. 

The membrane mechanics is expressed by the form of the stress tensor, $\boldsymbol{\sigma}$, and the force density, ${\bf f} = \nabla \cdot \boldsymbol{\sigma}$, acting on it~\cite{salbreux2017,sahu2017,torres-sanchez2019}. We describe this briefly in coordinate-free form. The force has normal and tangential components and, likewise, so does the stress. The tangential stresses are equivalent to a linear transformation on the tangent space and can be described in terms of the natural structure of these, as we illustrate in Fig.~\ref{fig:stresses}(a). They separate into isotropic and anisotropic parts, each of which forms a two-dimensional subspace. The isotropic subspace is spanned by the identity, ${\bf I}_2$, and the complex structure, ${\bf J}$~\cite{KarcherNotes,Nakahara,note_LeviCivita}, whose square is minus the identity, ${\bf J}^2 = - {\bf I}_2$, and which acts by $90^{\circ}$ rotation about the surface normal, ${\bf J}\cdot{\bf a} = {\bf N} \times {\bf a}$ for any tangent vector ${\bf a}$. The shape operator of the membrane provides a basis for the anisotropic subspace, through its decomposition ${\bf S} = H {\bf I}_2 + {\bf D}$ into isotropic and anisotropic parts. The pair ${\bf D}, {\bf J}\cdot{\bf D}$ then forms a basis for the anisotropic subspace~\cite{note_umbilic}. In the principal curvature basis they are represented by 
\begin{align}
    & {\bf D} = \frac{k_1-k_2}{2} \begin{bmatrix} 1 & 0 \\ 0 & -1 \end{bmatrix} , 
    && {\bf J}\cdot{\bf D} = \frac{k_1-k_2}{2} \begin{bmatrix} 0 & 1 \\ 1 & 0 \end{bmatrix} .
\end{align}
We illustrate the basic tangential stresses in Fig.~\ref{fig:stresses}(a). 

The normal stresses are of the form ${\bf a}\,{\bf N}$, where ${\bf a}$ is a tangent vector coming from the geometry. The natural choices are the gradients of the scalar curvatures, $\nabla H$, $\nabla K_{\textrm{G}}$; here we restrict to the gradient of mean curvature.  
We summarise in Table~\ref{tab:stress_forces} this basis of geometric membrane stresses and the force density associated to each of them. Other contributions are formed from these by composition, or multiplication by a function of the scalar curvatures. Details of calculating the surface divergences are given in the Supplemental Material (SM)~\cite{suppMat}. 

\begin{table}[b]
    \centering
    \begin{tabular}{|c|c|}
     \hline stress & force density \\
      \hline
      ${\bf I}_2$ & $2H\,{\bf N}$ \\
      ${\bf J}$ & $0$ \\
      ${\bf D}$ & $\nabla H + 2(H^2 - K_{\textrm{G}}) {\bf N}$ \\
      ${\bf J}\cdot{\bf D}$ & ${\bf J}\cdot\nabla H$ \\
      $\nabla H \,{\bf N}$ & $\nabla^2 H\,{\bf N} - H \nabla H - {\bf D}\cdot \nabla H$ \\
      ${\bf J}\cdot\nabla H \,{\bf N}$ & ${\bf J}\cdot{\bf D}\cdot\nabla H - H {\bf J}\cdot\nabla H$\\   \hline    
    \end{tabular}
    \caption{Basis of geometric membrane stresses and their associated force denisties. Other contributions are formed from these by composition or multiplication by a scalar.} 
    \label{tab:stress_forces}
\end{table}

We now describe some generic aspects of these stresses, starting with those of an equilibrium membrane. The geometric stresses of an equilibrium membrane are distinguished by having forces that are directed purely along the surface normal. The simplest of these are proportional to ${\bf I}_2$, ${\bf D}-H{\bf I}_2$ and $\nabla H \,{\bf N} + H{\bf D}$. Together, they produce the equilibrium membrane stress~\cite{capovilla2002} 
\begin{equation}
\begin{split}
    \boldsymbol{\sigma}^{\textrm{eq}} & = \gamma \,{\bf I}_2 - 2\kappa \bigl( \nabla H \,{\bf N} + H{\bf D} \bigr) \\
    & \quad + 2\kappa H_0 \bigl( {\bf D} - (H-H_0) {\bf I}_2 \bigr) ,
    \label{eq:equilibrium_stress}
\end{split}
\end{equation}
where $\gamma$ is the membrane tension, $\kappa$ is the bending modulus and $H_0$ is the spontaneous mean curvature. In general $\gamma$ can be a scalar function on the membrane, however, in the case where it is constant the equilibrium stresses are associated to a free energy 
\begin{equation}
    F = \int \biggl( \gamma + \frac{\kappa}{2} \bigl( 2H - 2H_0 \bigr)^2 \biggr) dA ,
    \label{eq:free_energy}
\end{equation}
with $dA$ denoting the surface area element. 

Conversely, any geometric stress that produces a non-zero tangential force is associated to a non-equilibrium process. For example, a stress proportional to ${\bf D}$ alone, and not ${\bf D}-H{\bf I}_2$, contributes a tangential force proportional to $\nabla H$ and acts like an active Marangoni-like stress~\cite{salbreux2017,mietke2019}. 
The basic stresses ${\bf J}$, ${\bf J}\cdot{\bf D}$ and ${\bf J}\cdot\nabla H \,{\bf N}$ all generate purely tangential forces. 
We refer to them as chiral stresses as they involve an odd power of the complex structure ${\bf J}$. They represent a geometric form of odd elasticity~\cite{scheibner2020,fruchart2023}, coupling elastic membrane bending to in-plane stresses. This differs from ordinary odd elasticity in both that the response is to bending deformations rather than in-plane strains and that the resulting force acts along a direction that is fluid rather than solid. As such, they are analogous to the odd elasticity of active cholesterics~\cite{kole2021} but here realised in a two-dimensional membrane.

Some insight into the active stresses can be gained from comparison with a thin film of chiral active gel~\cite{prost2015,julicher2018}, see Fig.~\ref{fig:stresses}(b). Such a gel with polarisation ${\bf p}$ has an achiral active stress $\propto {\bf pp}$, associated to microscopic force dipoles, and a chiral active stress $\propto \nabla\times({\bf pp}) + \bigl[ \nabla\times({\bf pp}) \bigr]^T$, associated to microscopic torque dipoles~\cite{furthauer2012,furthauer2013}. For a thin film polarised (uniformly) along its normal direction, ${\bf p} = {\bf N}$, we can write the achiral stress as ${\bf pp} = {\bf I}_3 - \bigl( {\bf I}_3 - {\bf NN} \bigr)$, where ${\bf I}_3$ is the identity in $\mathbb{R}^3$, and see that, modulo a bulk pressure in the film, it behaves as an active membrane tension~\cite{daRocha2022}. 
Similarly, for the chiral active stress we can write 
\begin{equation}
    \nabla \times \bigl( {\bf pp} \bigr) = \bigl( \nabla \times {\bf N} \bigr) {\bf N} - \bigl( {\bf N} \times \nabla \bigr) {\bf N} = - H {\bf J} + {\bf J}\cdot{\bf D} ,
\end{equation}
so that the chiral activity of the gel acts like a membrane stress proportional to ${\bf J}\cdot{\bf D}$. Thus such a stress arises naturally from the standard theories of active gels and liquid crystals with microscopic active torques.

For the remainder of this paper we restrict ourselves to the phenomenology of chiral stresses, for which we take the active stress 
\begin{equation}
    \boldsymbol{\sigma}^{\textrm{a}} = \zeta \,{\bf J}\cdot{\bf D} + \xi \,{\bf J}\cdot\nabla H \,{\bf N} ,
\label{eq:chiral_active_stress}
\end{equation}
retaining the lowest order (symmetric) tangential and normal contributions. This implicitly includes the chiral stress $H{\bf J}$ as any differences between this and ${\bf J}\cdot{\bf D}$ come only from effects at the boundary, which we do not discuss here. For a more detailed discussion of the equivalence of these stresses see SM~\cite{suppMat}.

\textbf{Viscous stress and dynamical equations: }
We write the membrane velocity as ${\bf V} = {\bf v} + \mathrm{v}_\mathrm{n} \,{\bf N}$, where ${\bf v}$ is the tangential component. Membrane incompressiblity is given by 
\begin{equation}
    \nabla \cdot {\bf V} = \nabla \cdot {\bf v} - 2H \mathrm{v}_\mathrm{n} = 0.
\label{eq:continuity}
\end{equation}
The velocity gradients can be decomposed into tangential and normal parts as $\nabla {\bf V} = \nabla_{\parallel} {\bf v} - \mathrm{v}_\mathrm{n} {\bf S} + \bigl( {\bf S}\cdot{\bf v} + \nabla \mathrm{v}_\mathrm{n} \bigr) {\bf N}$, where $\nabla_{\parallel} {\bf v} = \nabla {\bf v} - (\nabla {\bf v} \cdot {\bf N}){\bf N}$ is the covariant derivative of ${\bf v}$. The viscous stress comes from the symmetric part of the purely tangential velocity gradients 
\begin{equation}
    \boldsymbol{\sigma}^{\textrm{v}} = \eta \Bigl( \nabla_{\parallel} {\bf v} + (\nabla_{\parallel} {\bf v})^T - 2\mathrm{v}_\mathrm{n} {\bf S} \Bigr) ,
\label{eq:viscous_stress}
\end{equation}
where $\eta$ is the membrane shear viscosity \cite{scriven1960,arroyo2009}. For motions of viscous membranes at microscopic scales, inertia can be neglected and Newton's law reduces to the force balance $0 = \nabla \cdot (\boldsymbol{\sigma}^{\textrm{eq}} + \boldsymbol{\sigma}^{\textrm{v}} + \boldsymbol{\sigma}^{\textrm{a}})$, which reads (for vanishing spontaneous curvature $H_0=0$)
\begin{equation}
\begin{split}
& \eta \Bigl[ \nabla_{\parallel}^2 {\bf v} + K_{\textrm{G}} \,{\bf v} - 2 \mathrm{v}_{\textrm{n}} \nabla H  - 2 {\bf D} \cdot \nabla \mathrm{v}_{\textrm{n}} \Bigr] \\
& +\nabla \gamma + \zeta \,{\bf J}\cdot\nabla H - \xi {\bf S}\cdot{\bf J}\cdot\nabla H  \\
& + \Bigl[ 2\gamma H - 2\kappa \bigl( \nabla^2 H + 2H (H^2 - K_{\textrm{G}}) \bigr) \\
& \quad + 2\eta \bigl( \nabla_{\parallel} {\bf v} : {\bf D} - 2\mathrm{v}_\mathrm{n} (H^2 - K_{\textrm{G}}) \bigr) \Bigr] {\bf N} = 0.
\end{split} 
\label{eq:force_balance}
\end{equation}
Here, the spatially varying surface tension can be viewed as a Lagrange multiplier imposing the surface incompressibility condition, as such it is essentially the negative of a two-dimensional pressure on the surface \cite{arroyo2009,rangamani2013}. If the membrane is closed a pressure difference term should be added to the normal force. Achiral active stresses would contribute as per Table~\ref{tab:stress_forces}. 
Equations~\eqref{eq:continuity} and~\eqref{eq:force_balance} provide the governing equations for the membrane in coordinate-free form. They are equivalent, with some simplification, to those obtained in Refs.~\cite{salbreux2017,sahu2017} but the presentation in terms of intrinsically-defined quantities is complementary and the phenomenology of chiral activity is unexplored. We note that it is possible to include other generalised viscosities, such as a bending viscosity~\cite{salbreux2022}, however we do not discuss such terms here.

\begin{figure*}[t!]
\centering
\includegraphics[width=\textwidth]{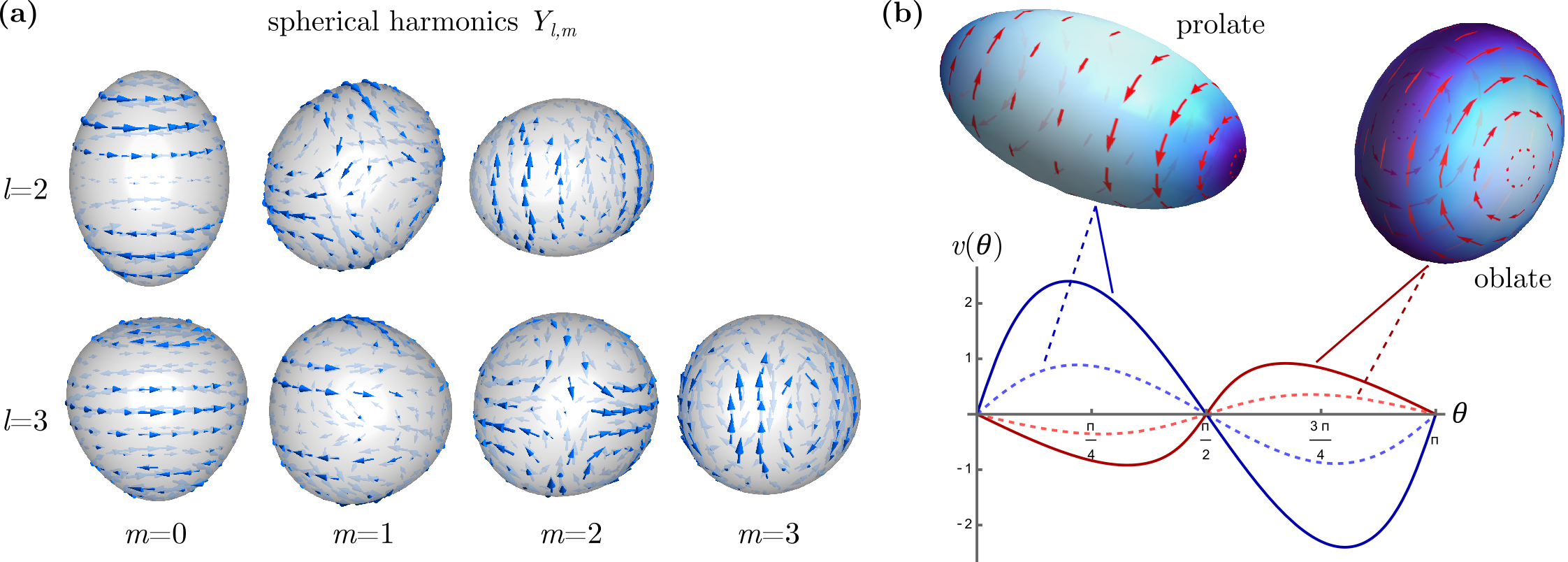}
\caption{\textbf{Odd mechanics of a sphere and spontaneous flow on a spheroid.} \textbf{(a)} Odd mechanics of low order multipole deformations of a sphere expressed in terms of real spherical harmonics $Y_{l,m}$. The arrows denote the flow induced by the odd force from gradients in mean curvature. \textbf{(b)} Flows on a spheroid where the colour indicates the mean curvature. Graph shows azimuthal velocity, $v(\theta)$, for both prolate and oblate spheroids. Note the switch in handedness of flows from prolate to oblate.}
\label{fig:spheroidFlows}
\end{figure*}

The chiral activity enters directly only into the tangential force balance, where it serves to drive membrane flows. It then also influences the membrane shape through the normal component of the viscous force, or Scriven-Love term~\cite{scriven1960,sahu2020}. The relative strength of this force as compared to membrane bending forces is described by the dimensionless Scriven-Love number, $\mathrm{SL} = \eta V L/\kappa$ \cite{sahu2020,al-izzi2023}, where $V$ is a characteristic velocity and $L$ a lengthscale. For the active chiral stresses the velocity scale is $V\sim \zeta/\eta$ or $V\sim \xi /\eta L$, giving Scriven-Love numbers (or dimensionless activities) $\bar{\zeta} = \zeta L / \kappa$ and $\bar{\xi} = \xi / \kappa$. For sufficiently large values of either the normal viscous force can overcome membrane tension and bending, allowing for a novel membrane instability where perturbations to a non-flowing steady state create chiral flows whose Scriven-Love force enhances the perturbation. For this, the velocity gradients $\nabla_{\parallel} {\bf v}$ need to contain a part proportional to ${\bf D}$, which can come either from an extensional flow aligned with the principal curvature axes or from a shear flow at $\pm 45^{\circ}$ to them, see Fig.~\ref{fig:stresses}(c) for a schematic illustration. We give explicit examples below of this odd mechanical instability for membrane tubes and helicoids. 

It is worth noting that the active chiral stresses have a crossover lengthscale associated with them, $L_\mathrm{c}=\xi/\zeta$. For lengthscales less than $L_\mathrm{c}$ the active stress $\sim {\bf J}\cdot{\bf D}$ dominates the mechanics giving rise to forces along lines of constant mean curvature whose magnitude is set by the perpendicular gradient in mean curvature. For larger lengthscales the mechanics is dominated by the stress $\sim{\bf J}\cdot\nabla H{\bf N}$ which has a similar mechanism but now with components weighted by the principal curvatures. For our examples we will consider only the first case and set $\xi=0$ from this point onwards.

\textbf{Odd mechanics and counter-rotating flows in spheroidal membranes:} 
We now describe more explicitly the odd mechanics for a spherical membrane; a similar analysis can also be given for planar membranes. A sphere of radius $R$ is an exact solution of~\eqref{eq:continuity} and~\eqref{eq:force_balance}, with a pressure difference $\Delta P = 2\gamma/R$, that is linearly stable to the chiral activity since for a sphere ${\bf D} = 0$ and the Scriven-Love term vanishes to linear order. We consider an imposed normal displacement $\psi$ and determine the membrane flow to linear order from the tangential force balance. It is natural to analyse the perturbations in terms of spherical harmonics, $Y_{l,m}$. A uniform change in radius ($l=0$) does not induce any tangential flow, nor does a dipole mode ($l=1$) as to linear order it represents only a translation. Higher harmonics induce a tangential flow~\cite{suppMat} 
\begin{equation}
{\bf v} = - \frac{\zeta}{2\eta} \,{\bf J} \cdot \nabla \psi \text{,}
\label{eq:odd_sphere_flow}
\end{equation}
proportional to the perturbation gradient but directed perpendicularly to it. For positive $\zeta$ the flow is a right-handed circulation around maxima in the displacement. In Fig.~\ref{fig:spheroidFlows}(a) we show the form of the response for spherical harmonics at quadrupole ($l=2$) and octupole ($l=3$) order. $\psi$ can be viewed as an Eulerian displacement field for the membrane, so that the response~\eqref{eq:odd_sphere_flow} has the character of odd elasticity in being linear in deformation gradients with a twist~\cite{scheibner2020,fruchart2023}.

The simplest deformation of the sphere is an axisymmetric stretch to a spheroidal shape. This geometry is also amenable to an exact solution for the tangential flows, beyond the linear response described in the previous paragraph, allowing the full nonlinear geometry to be explored. For a spheroid with principal semi-axes $a,a,c$ described by the embedding ${\bf X} = (a \sin\theta \cos\phi , a \sin\theta \sin\phi , c \cos\theta)$, we obtain the exact azimuthal flow ${\bf v} = v(\theta) \,{\bf e}_{\phi}$ in the SM~\cite{suppMat}. We plot examples of the flows for both prolate ($c>a$) and oblate ($c<a$) spheroids in Fig.~\ref{fig:spheroidFlows}(b). To first order in $( \frac{c}{a} - 1 )$ the azimuthal velocity is 
\begin{equation}
    v(\theta) = \frac{\zeta}{\eta} \biggl( \frac{c}{a} - 1 \biggr) \sin\theta \cos\theta \mathrm{,}
\end{equation} 
reproducing the general result~\eqref{eq:odd_sphere_flow}. 
Odd mechanics generates circulating flows in the two `hemispheres' along the stretch axis that are right-handed, with $\zeta > 0$, for a prolate spheroid and left-handed for an oblate spheroid. In the solution to the full non-linear geometry one finds that the position of maximum velocity is that of the highest gradient in mean curvature. That is, offset closer to the poles in the case of a prolate spheroid and closer to the equator for an oblate spheroid, see Fig.~\ref{fig:spheroidFlows} (b) and \cite{suppMat} for details.

Intriguingly, exactly such flows are observed in {\it C. elegans} zygotes as they undergo left-right symmetry breaking before the principal division~\cite{naganathan2014}. In Ref.~\cite{naganathan2014} chiral active gel theory theory was used to account for these flows by assuming a varying strength of activity over a thin layer of cortex. Our results establish that the same flows can be obtained from uniform chiral activity purely as a consequence of the geometry. The geometric mechanism predicts that the direction of circulation reverses between prolate and oblate spheroids, which could be tested using experiments on reconstituted active chiral membranes confined within artificial eggshells of varying geometry.  

\textbf{Instability of a membrane tube:} For any surface which is not isotropic (totally umbilic), small surface deformations will drive flows that, in turn, couple back into the shape equation by the Scriven-Love force. If these forces are sufficiently large they can drive instabilities in the membrane shape. To illustrate this with a concrete example we take the case of a membrane tube. Such structures occur readily in biology, from the membrane tubes of the endoplasmic reticulum~\cite{nixon-abell2016}  to tissue structures such as the \textit{Drosophila} hindgut \cite{taniguchi2011,inaki2018}. 

\begin{figure}[t]
    \centering
   \includegraphics[width=8.6cm]{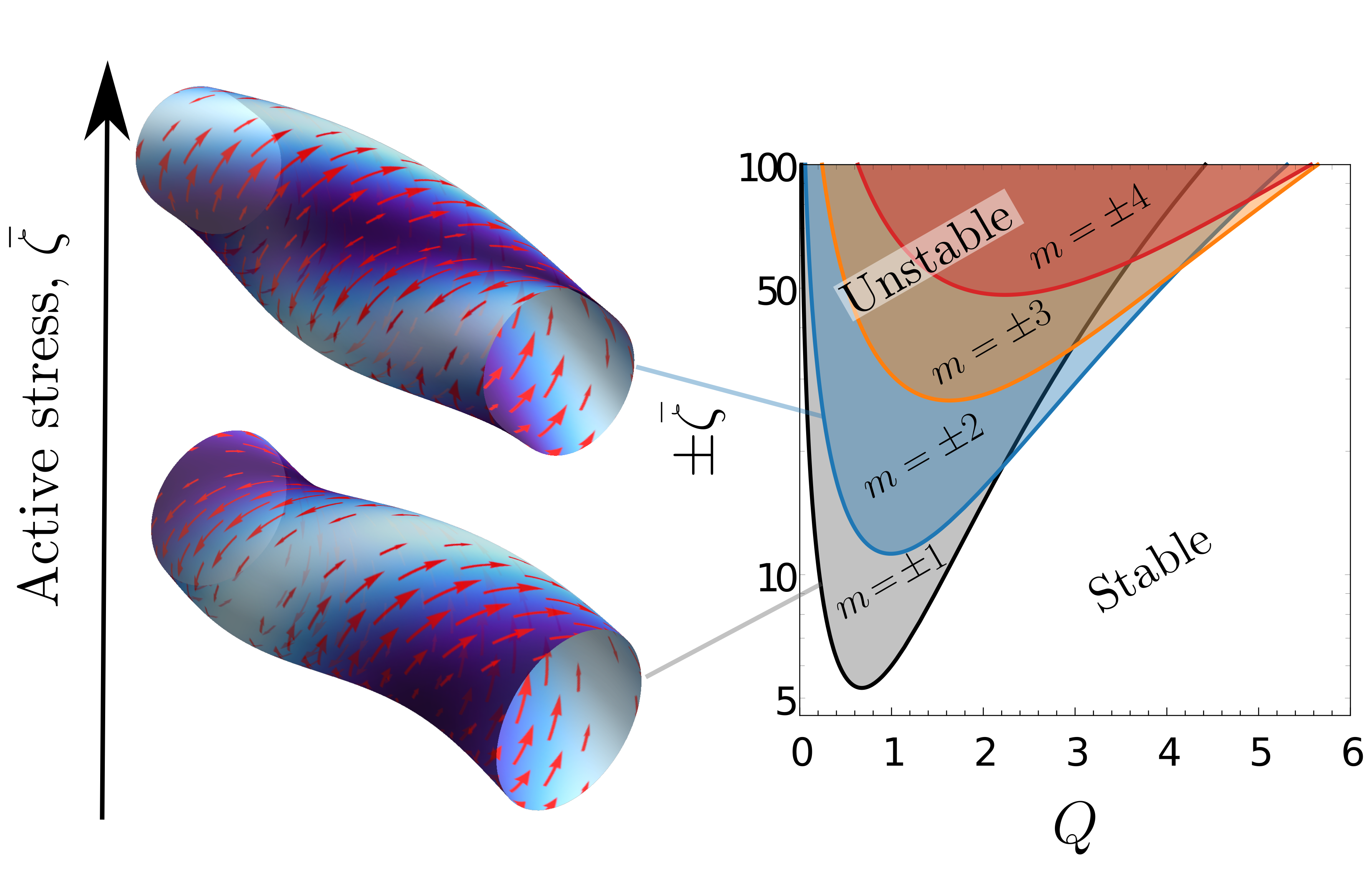}
    \caption{\textbf{Instability of an odd active membrane tube.} Stability diagram of wavenumber along the tube, $Q$, against active chiral stress (Scriven-Love number), $\bar\zeta$, for $m=\pm 1,2,3,4$. Schematic of the first two left-handed unstable modes ($m=1,2$) with mean curvature coloured on the surface and vectors showing the flow along lines of constant mean curvature.}
    \label{fig:oddActiveInstability}
\end{figure}

The principal curvature directions on the tube are axial and azimuthal so that a diagonal shear flow, winding in barber-pole fashion, will generate a normal Scriven-Love force. When the shape change this produces amplifies the shear flow, the tube will undergo a linear instability. This is shown in Fig.~\ref{fig:oddActiveInstability}. 
We consider perturbations of the tube radius by an amount $\psi_{m}(q,t) \,\mathrm{exp}\{i(m\theta+qz)\}$ about its passive equilibrium radius $r=\sqrt{\kappa/(2\gamma)}$, where $\theta,z$ are azimuthal and axial coordinates and $\psi_{m}/r \ll 1$. The general equations~\eqref{eq:continuity} and~\eqref{eq:force_balance} give a growth equation for the mode $\partial_t \psi_m \sim G_m(q) \psi_{m}$ (see SM for details~\cite{suppMat}). Nondimensionalising with lengthscale $r$, timescale $\tau=\eta r^2/\kappa$ and defining the active chiral Scriven-Love number as $\bar{\zeta}=r\zeta/\kappa$ we find the dimensionless growth rate as
\begin{equation}
    \begin{split}
        &G_m = \frac{\left(m^2+Q^2\right)}{4 Q^4} \Big[\bar\zeta m Q \left(m^2+Q^2-1\right) \\
        &- \left(m^2+Q^2\right) \left(m^4+2 m^2 \left(Q^2-1\right) +Q^4+1\right)\Big] \text{.}
    \end{split}
\label{eq:tube_growth}
\end{equation} 
For large enough values of $\bar{\zeta}$ the membrane tube is linearly unstable; the lowest threshold is for the $|m|=1$ mode with threshold at $\bar\zeta \approx 5.3$. Crucially for a mode to be unstable it must have $\bar\zeta m Q>0$, which explicitly selects a chirality to the shape instability. That is, for positive $\bar{\zeta}$, only modes with $mQ>0$ go unstable, \textit{i.e.}~left handed helical deformations. The dimensionless growth rate $G_m$ is shown in the SM \cite{suppMat} and the stability diagram is shown in Fig.~\ref{fig:oddActiveInstability}. In the $|m|=1$ case, the peak wavelength is given by $Q^*\approx 4/\bar\zeta$ in the small $Q$ regime so that the wavelength of the modulation is comparable to the tube circumference. For the $|m|=1$ shape instability note that flows are driven exactly along the lines of active force, that is along lines of constant mean curvature, Fig.~\ref{fig:oddActiveInstability}. The helical flows generated here are similar in nature to the flows on a cylinder of fixed geometry with active nematic stress~\cite{napoli2020}. 
As the magnitude of the activity is increased the most unstable mode crosses over to higher $|m|$, $|m|=2, 3, \dots$, so that different morphologies can be realised at sufficiently large activity. The crossover of fastest growing mode from $|m|=1$ to $|m|=2$ occurs at $\bar\zeta\approx 12.4$. 
Note that in the $|m|=2$ case the flows do not correspond perfectly with lines of constant mean curvature as the passive viscous forces which act to relax the shape of the tube (\textit{i.e.}~ aligned with gradients in mean curvature) play a larger role in the dynamics than in the case of the helical bending mode. 

Due to the ubiquitous nature of membrane tubes across many scales in biology, such a generic odd elastic instability could play a role in a variety of different systems. Of particular note is the morphogenesis of the \textit{Drosophila} hindgut, where an epithelial tissue tubule undergoes a chiral twisting motion to form a characteristic `question mark' shape \cite{taniguchi2011,inaki2018}. We speculate that such an odd elastic instability may be the driving force behind this shape transition.

\textbf{Instabilities of minimal surfaces:} As an additional illustration of these generic Scriven-Love type instabilities we turn to the case of minimal surfaces. Such surfaces have been considered extensively as examples of necks \cite{chabanon2018,fonda2021} and helical ramps in lipid membranes \cite{chabanon2019,teresaki2013,guven2014}. In addition such surfaces have generated considerable interest as they describe the geometry of soap films \cite{boudaoud1999,machon2016PRL,alexander2020,powers2002,goldstein2014}.

\begin{figure*}[t!]
    \centering
   \includegraphics[width=\textwidth]{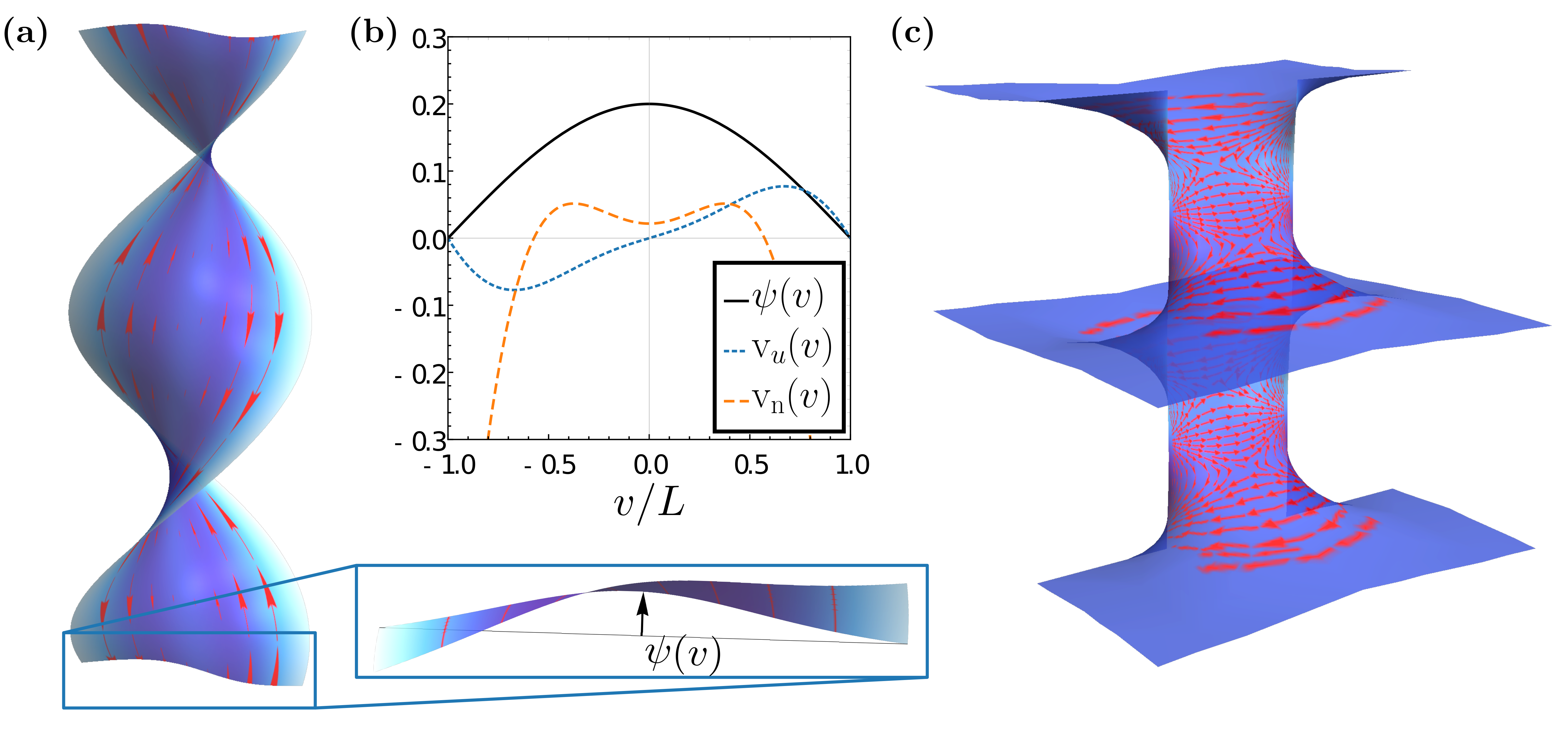}
    \caption{\textbf{Chiral flows and instabilities of active helical structures.} \textbf{(a)} Schematic of the helicoid with first nodal perturbation with mean curvature coloured on the surface and vectors showing the flow along lines of constant mean curvature. Inset gives a zoom in to the perturbation, as viewed from above. \textbf{(b)} First nodal perturbation $\psi(v) = \delta\cos\left(\pi v/2 L\right)$, corresponding spontaneous flow field $\mathrm{v}_u(v)$ and normal velocity $\mathrm{v}_\mathrm{n}(v)$ plotted for $\delta=0.2$, $\tilde\zeta=-10$ and $\alpha L =1$. \textbf{(c)} Streamline surface plot of the chiral force ${\bf J}\cdot\nabla H$ on a ``parking garage'' approximately minimal surface formed from the addition of helicoidal ramps for separation $R=2$, and pitch $p_0=1$.}
    \label{fig:oddHelicoid}
\end{figure*}

We focus on the example of a helicoid, given by the embedding ${\bf X}_{\text{helicoid}}(u,v) = \left( \sinh\left(\alpha v\right)\cos\left(\alpha u\right)/\alpha  , \sinh\left(\alpha v\right)\sin\left(\alpha u\right)/\alpha , u \right)$, where $\alpha$ is the wavenumber of the helicoid. The shape operator has purely off-diagonal form  
\begin{equation}\label{eq:helicoidShapeOpUnperturbed}
   {\bf S} = - \nabla {\bf N} = \alpha \,\textrm{sech}^2 \alpha v \bigl( {\bf e}_u {\bf e}_v + {\bf e}_v {\bf e}_u \bigr)\text{,} 
\end{equation}
which couples to a non-zero Scriven-Love force for shear flows directed along the helicoid axis. It is the instability associated to this that we analyse. Perturbations of the helicoid in the normal direction that vary along the transverse $v$ coordinate will drive spontaneous shear flows along the $u$ direction (helicoid axis) and their coupling to the normal Scriven-Love force creates the shape instability; see Fig.~\ref{fig:oddHelicoid} (a) for an illustration of this phenomenology. 

Here we consider a finite strip $v\in [-L,L]$ of the helicoid and the lowest order perturbation of the form ${\bf X}_{\text{helicoid}} + \delta  \cos(\pi v/2L){\bf N}$. 
For simplicity we set $\kappa=0$ here, which will alter the shape dynamics, but leave the instability criterion unchanged. One component of the tangential force balance and the continuity equation gives $\gamma(v)=\gamma_0$ and ${\bf v}\cdot{\bf e}_v=0$, which then leaves a system of coupled non-linear boundary value ODEs for the normal velocity ($\mathrm{v}_\text{n}$) and tangential velocity ($\mathrm{v}_u={\bf v}\cdot{\bf e}_u$). In order to make progress we neglect forces from the normal velocity in the tangential force balance and compute the flows, which we show in Fig.~\ref{fig:oddHelicoid}(a) \& (b). We then compute the normal velocity $\mathrm{v}_\mathrm{n}(v)$, Fig.~\ref{fig:oddHelicoid}(b), allowing us to find conditions for which parts of the initial perturbation grow; the region of positive normal velocity in Fig.~\ref{fig:oddHelicoid}(b). For full derivation see SM~\cite{suppMat}. The normal perturbations can be driven unstable when the Scriven-Love force becomes larger than the restoring surface tension. Crucially this occurs only when the active stress $\zeta$ and helical wavenumber $\alpha$ have opposite signs, giving a handedness to the instability. This is perhaps to be expected as the chirality of the helicoid essentially selects a handedness for the instability. Numerically we find the threshold criterion $\zeta\alpha\lessapprox-6\gamma$. 

Finally, we consider an approximately minimal surface formed by the addition of two helicoidal ramps of opposite pitch. In complex coordinates we write this as $X(z)=(z,p_0[\arg(z-R)-\arg(z+R)])$ where $p_0$ is the magnitude of the pitch and $2R$ is the separation. This type of configuration has been used to discuss the morphology of helical ``parking garage'' ramps seen in several organelles such as the rough endoplasmic reticulum (ER) \cite{teresaki2013,guven2014,bussi2019}. Although it is possible to map such surfaces onto exact minimal surfaces \cite{dasilva2021}, here we consider the approximate surfaces as a perturbation and examine the forces that arise from $H\neq 0$. We plot the streamlines of the force ${\bf J}\cdot\nabla H$ in Fig.~\ref{fig:oddHelicoid}(c), which shows vortex/anti-vortex sheets on the ramps of the structure, see SM for details \cite{suppMat}. We speculate that chirally actuating proteins, \textit{e.g.}~the ATP synthase \cite{nirody2020}, could induce such forces thereby enabling increased mixing between the layers of organelles due to Taylor dispersion. Such active mixing could have significant implications for protein and lipid production in the ER.

\textbf{Discussion:} In this paper we have presented coordinate-free equations for membrane hydrodynamics with active chiral stresses. Such a presentation has allowed us to identify clear geometric phenomenology associated with these active stresses. This includes the generation of forces which are proportional to gradients in mean curvature and act in directions of constant mean curvature, or weighted by the principal curvature directions in the two simplest cases respectively. These stresses represent a novel form of odd elasticity characterised by normal deformations giving rise to asymmetric tangential forces. Such stresses lead to spontaneous shear flows in the presence of mean curvature gradients, and shape instabilities characterised by an active counterpart of the usual Scriven-Love number. To illustrate the generic nature of the flows and shape instabilities we have given explicit examples for spheroids, tubes and helicoids.

We note that many of these examples bear a resemblance to phenomena seen in biology, including the counter-rotating flows of the \textit{C. elegans} zygote, and the \textit{Drosophila melanogaster} hindgut. Here our geometric representation gives clear benefits as the flows along lines of constant mean curvature could be used to infer the type of active stress in tissues using modern $4$D microscopy techniques, such as lattice light sheet microscopy \cite{chen2014,serra2020}. 

There remain significant open challenges associated with numerically solving, in full generality, the types of morphodynamic equations discussed in this paper, although recent work developing stable finite element schemes for lipid membrane hydrodynamics should be extendable to such active chiral systems \cite{torres-sanchez2019,sahu2020b,barrett2016}. Adapting such methods to account for chiral stresses will allow for studies of increasingly realistic biological and soft matter systems and open the door to simulating increasingly rich theories of quantitative mechanobiology.

\acknowledgements{SCA-I \& GPA thank Nordita for funding and hospitality during the Current and Future Themes in Soft \& Biological Active Matter program. SCA-I acknowledges funding the EMBL-Australia program. The authors acknowledge beneficial discussions with Matthew Turner, Andreas Carlson, Diana Khoromskaia, Jack Binysh and Richard Morris.}

\include{supplement.tex}

\end{document}

%% file: supplement.tex
%









\clearpage
\onecolumngrid
\makeatletter 
\def\tagform@#1{\maketag@@@{(S\ignorespaces#1\unskip\@@italiccorr)}}
\makeatother
\graphicspath{{figures/}} 
\makeatletter
\makeatletter \renewcommand{\fnum@figure}
{\figurename~S\thefigure}
\makeatother
\def\eq#1{{Eq.~S\ref{#1}}}    
\def\fig#1{{Fig.~S\ref{#1}}}
\setcounter{figure}{0} 
\setcounter{equation}{0} 
\appendix
\section*{Supplemental Material for ``A Twist On Active Membranes: Odd Mechanics, Spontaneous Flows and Shape Instabilities''}\label{sec:SI}
\begin{center}
    Sami C.~Al-Izzi \& Gareth P.~Alexander
\end{center}

This supplement consists of three parts, each divided in multiple subsections. In the first part we provide a short summary of basic notions of differential geometry and establish our notational conventions. In the second part we collate calculations of the geometric and hydrodynamic membrane forces. The third part is the longest and presents, in order, further calculational details for the examples of spheroids, membrane tubes and minimal surfaces that we described in the main text. 

\subsection{I. Elements of differential geometry and notational conventions}

\subsubsection{Membrane geometry}
\label{SI:geometry}

In this section we give a short summary of the geometry of surfaces using the shape operator, covering those aspects that we make use of in the main text. Some general references include~\cite{doCarmo,Kobayashi,GrayAbbenaSalamon,KarcherNotes}. Our style of presentation is similar to analogous analysis of director gradients in liquid crystals~\cite{machon2016PRX,machon2019,selinger2018}.

We consider the membrane to be a surface smoothly embedded in $\mathbb{R}^3$ and represent its points, within $\mathbb{R}^3$, by ${\bf X}$. We denote the surface normal by ${\bf N}$. Derivatives of quantities on the membrane are induced from those of the ambient space. The surface gradients are the gradients of $\mathbb{R}^3$ with the restriction that only derivatives along tangential directions are possible; in an abuse of notation we denote this $\nabla_{{\bf N}} \equiv 0$. We represent the geometry of the membrane using the shape operator ${\bf S} = -\nabla {\bf N}$. This is a linear transformation on the tangent space of the surface and we decompose it with respect to the natural structure for these. There are two canonical linear transformations, the identity ${\bf I}_2$ and the complex structure ${\bf J}$, which are both isotropic with respect to the action of rotations about the surface normal. In any (positive) orthonormal basis they are represented by the matrices 
\begin{align}
    & {\bf I}_2 = \begin{bmatrix} 1 & 0 \\ 0 & 1 \end{bmatrix} , 
    && {\bf J} = \begin{bmatrix} 0 & -1 \\ 1 & 0 \end{bmatrix} .
\end{align}

The remaining linear transformations are anisotropic and form a two-dimensional subspace, the elements of which transform amongst themselves under rotations about ${\bf N}$ as a spin 2 object. The shape operator provides a natural basis for the anisotropic subspace through its decomposition into isotropic and anisotropic parts  
\begin{equation}
{\bf S} = - \nabla {\bf N} = H {\bf I}_2 + {\bf D} .
\end{equation}
Here $H$ is the mean curvature and ${\bf D}$ lies in the subspace of anisotropic linear transformations; there is no component along ${\bf J}$ by the Frobenius integrability theorem. The eigenvectors of the shape operator ${\bf E}_1$, ${\bf E}_2$ are the principal curvature directions of the surface and with respect to this orthonormal basis the shape operator is represented by the matrix 
\begin{equation}
  {\bf S} = \frac{k_1 + k_2}{2} \begin{bmatrix} 1 & 0 \\ 0 & 1 \end{bmatrix} + \frac{k_1 - k_2}{2} \begin{bmatrix} 1 & 0 \\ 0 & -1 \end{bmatrix} ,
\end{equation}
where $k_1 > k_2$ are the principal curvatures. The mean curvature is $H = \frac{1}{2} (k_1+k_2)$ and $k_1 - k_2 = 2 \sqrt{H^2 - K_{\textrm{G}}}$, where $K_{\textrm{G}} = k_1 k_2$ is the Gaussian curvature. 

Composition of ${\bf D}$ with the complex structure defines a second anisotropic linear transformation, which we denote by ${\bf J}\cdot{\bf D}$. In the principal curvature basis it is represented by the matrix 
\begin{equation}
  {\bf J}\cdot{\bf D} = \frac{k_1 - k_2}{2} \begin{bmatrix} 0 & 1 \\ 1 & 0 \end{bmatrix} .
\end{equation}
With respect to the standard inner product on matrices ${\bf D}$ and ${\bf J}\cdot{\bf D}$ are orthogonal; they provide a basis for the anisotropic linear transformations on the tangent space. The eigenvectors of ${\bf J}\cdot{\bf D}$ are the directions ${\bf E}_{\pm} = \frac{1}{\sqrt{2}} ({\bf E}_1 \pm {\bf E}_2)$, making an angle $\pm \frac{\pi}{4}$ with the principal curvature directions. They may be called directions of `principal torsion'~\cite{machon2016PRX} as for curves in the surface along these directions the Darboux torsion is extremal.

\subsubsection{Normal variations}

We record here general first variation formulae for displacements of the surface along its normal~\cite{capovilla2002,capovilla2003,arroyo2009}. A normal variation of the surface is given by displacing the points according to ${\bf X} \mapsto {\bf X} + \psi \,{\bf N}$. To first order, the change in the surface normal is ${\bf N} \mapsto {\bf N} - \nabla \psi$. 
The first order variation of the shape operator is
\begin{equation}
    {\bf S} = - \nabla {\bf N} \mapsto {\bf S} + \nabla \nabla \psi + \psi \,{\bf S}\cdot{\bf S} + {\bf N} \bigl( {\bf S}\cdot\nabla \psi \bigr) + O(\psi^2) ,
    \label{SI:eq:S_variation}
\end{equation}
from which we can deduce the first order variation of the mean curvature  
\begin{equation}
    H \mapsto H + \frac{1}{2} \nabla^2 \psi + \psi \bigl( 2H^2 - K_{\textrm{G}} \bigr) + O(\psi^2) .
    \label{SI:eq:H_variation}
\end{equation}
The first order variation of ${\bf D}$ also follows from that for the shape operator 
\begin{equation}
    {\bf D} \mapsto \bigl( 1 + 2H \psi \bigr) \,{\bf D} + \nabla \nabla \psi - \frac{1}{2} \nabla^2 \psi \,{\bf I}_2 + {\bf N} \bigl( {\bf D}\cdot\nabla \psi \bigr) - H \nabla \psi \,{\bf N} + O(\psi^2) . 
    \label{SI:eq:D_variation}
\end{equation}
Finally, using $K_{\textrm{G}} = H^2 - \frac{1}{2} {\bf D}:{\bf D}$ we obtain 
\begin{equation}
    K_{\textrm{G}} \mapsto \bigl( 1 + 2H \psi \bigr) K_{\textrm{G}} + H \nabla^2 \psi - {\bf D} : \nabla\nabla\psi + O(\psi^2) .
    \label{SI:eq:K_variation}
\end{equation}

\subsection{II. Membrane forces}

\subsubsection{Geometric forces}

To compute the membrane forces we need to know about gradients of the shape operator. To summarise the basic identities~\cite{doCarmo,Kobayashi,GrayAbbenaSalamon} we make use of direct calculations in the principal curvature basis. We write the connection 
\begin{align}
    & \nabla {\bf E}_{1} = {\bf A} \,{\bf E}_2 + k_{1} \,{\bf E}_1 {\bf N} , 
    && \nabla {\bf E}_{2} = - {\bf A} \,{\bf E}_1 + k_{2} \,{\bf E}_2 {\bf N} ,
\end{align}
where ${\bf A}$ is the (dual to the) connection form on the tangent space. With this notation the gradient of the shape operator reads 
\begin{equation}
    \nabla {\bf S} = \nabla k_1 \,{\bf E}_1 {\bf E}_1 + \nabla k_2 \,{\bf E}_2 {\bf E}_2 + \bigl( k_1 - k_2 \bigr) {\bf A} \bigl( {\bf E}_1 {\bf E}_2 + {\bf E}_2 {\bf E}_1 \bigr) + k_1^2 \bigl( {\bf E}_1 {\bf E}_1 {\bf N} + {\bf E}_1 {\bf N} {\bf E}_1 \bigr) + k_2^2 \bigl( {\bf E}_2 {\bf E}_2 {\bf N} + {\bf E}_2 {\bf N} {\bf E}_2 \bigr) .
    \label{SI:eq:grad_shape}
\end{equation}
The Codazzi-Mainardi-Peterson equations~\cite{doCarmo,Kobayashi,GrayAbbenaSalamon} assert that ${\bf J}: \nabla {\bf S} = 0$, where the contraction is on the first two slots. This gives 
\begin{equation}
    0 = \bigl( {\bf E}_2 \cdot \nabla k_1 \bigr) {\bf E}_1 - \bigl( {\bf E}_1 \cdot \nabla k_2 \bigr) {\bf E}_2 + \bigl( k_1 - k_2 \bigr) \Bigl[ \bigl( {\bf E}_2 \cdot {\bf A} \bigr) {\bf E}_2 - \bigl( {\bf E}_1 \cdot {\bf A} \bigr) {\bf E}_1 \Bigr] ,
\end{equation}
and we find the connection is given by  
\begin{equation}
    \bigl( k_1 - k_2 \bigr) {\bf A} = \bigl( {\bf E}_2 \cdot \nabla k_1 \bigr) {\bf E}_1 + \bigl( {\bf E}_1 \cdot \nabla k_2 \bigr) {\bf E}_2 .
    \label{SI:eq:connectionA}
\end{equation}
We remark that this can be expressed in fully intrinsic form as 
\begin{equation}
    {\bf A} = \frac{1}{2(H^2-K_{\textrm{G}})} \,{\bf J} \cdot {\bf D} \cdot \nabla H - \frac{1}{4(H^2-K_{\textrm{G}})} \,{\bf J} \cdot \nabla \bigl( H^2-K_{\textrm{G}} \bigr) \text{.}
\end{equation}
Using \eqref{SI:eq:connectionA} we calculate the divergence of the shape operator to be  
\begin{equation}
    \nabla \cdot {\bf S} = 2 \nabla H + 2 \bigl( 2H^2 - K_{\textrm{G}} \bigr) {\bf N} ,
    \label{SI:eq:div_shape}
\end{equation}
and then decomposing ${\bf S}$ into its isotropic and anisotropic parts we obtain 
\begin{equation}
    \nabla \cdot {\bf D} = \nabla H + 2 \bigl( H^2 - K_{\textrm{G}} \bigr) {\bf N} .
    \label{SI:eq:div_D}
\end{equation}
For the tangent space identity ${\bf I}_2$ and complex structure ${\bf J}$ we record the short direct computations 
\begin{gather}
    \nabla \cdot {\bf I}_2 = \nabla \cdot \bigl( {\bf I}_3 - {\bf NN} \bigr) = - \bigl( \nabla \cdot {\bf N} \bigr) {\bf N} = \bigl( \textrm{tr} \,{\bf S} \bigr) {\bf N} = 2H \,{\bf N} , \\
    \nabla \cdot {\bf J} = \nabla \cdot \bigl( {\bf E}_2 {\bf E}_1 - {\bf E}_1 {\bf E}_2 \bigr) = \bigl( - {\bf A} \cdot {\bf E}_1 \bigr) {\bf E}_1 + \bigl( {\bf E}_2 \cdot {\bf A} \bigr) {\bf E}_2 - \bigl( {\bf E}_2 \cdot {\bf A} \bigr) {\bf E}_2 - \bigl( - {\bf A} \cdot {\bf E}_1 \bigr) {\bf E}_1 = 0 .    
\end{gather}
Finally, the divergence of ${\bf J}\cdot{\bf D}$ follows from writing it in the principal curvature basis, ${\bf J}\cdot{\bf D} = \frac{k_1-k_2}{2} \bigl( {\bf E}_1 {\bf E}_2 + {\bf E}_2 {\bf E}_1 \bigr)$, and then a short direct calculation 
\begin{align}
\begin{split}
    \nabla \cdot \bigl( {\bf J} \cdot {\bf D} \bigr) & = \frac{1}{2} \nabla (k_1 - k_2) \cdot \bigl( {\bf E}_1 {\bf E}_2 + {\bf E}_2 {\bf E}_1 \bigr) + \frac{k_1-k_2}{2} \Bigl[ \bigl( -{\bf A}\cdot{\bf E}_1 \bigr) {\bf E}_1 + \bigl( {\bf E}_2\cdot{\bf A} \bigr) {\bf E}_2 \\
    & \qquad + \bigl( {\bf E}_2\cdot{\bf A} \bigr) {\bf E}_2 + \bigl( -{\bf A}\cdot{\bf E}_1 \bigr) {\bf E}_1 \Bigr] , 
\end{split} \\
    & = \frac{1}{2} \Bigl[ -{\bf E}_2\cdot\nabla (k_1+k_2) {\bf E}_1 + {\bf E}_1\cdot\nabla (k_1+k_2) {\bf E}_2 \Bigr] = {\bf J}\cdot\nabla H \text{.}
\end{align}
This summarises the calculation of the membrane forces as recorded in Table~\ref{tab:stress_forces} of the main text.

\subsubsection{Equivalence of stresses \& the Belinfante-Rosenfeld tensor}

Certain stresses are divergence-free and hence produce no bulk force; we refer to them as null stresses. We are unaware of a general characterisation, however, the simplest examples are ${\bf J}$, $H{\bf J} + {\bf J}\cdot{\bf D}$ and $H^2{\bf J} + H{\bf J}\cdot{\bf D} - {\bf J}\cdot\nabla H \,{\bf N}$. As an interpretation, the second of these establishes the equivalence, at the level of forces, of the antisymmetric tangential stress $H{\bf J}$ and the symmetric tangential stress $-{\bf J}\cdot{\bf D}$. Further, it can be written as the divergence of a third rank tensor, $H{\bf J}+{\bf J}\cdot{\bf D} = \nabla \cdot ({\bf J} {\bf N})$, and in this sense null stresses are reminiscent of the Belinfante-Rosenfeld tensor \cite{belinfante1940,rosenfeld1940,gotay1992}. We remark that $H^2{\bf J}+H{\bf J}\cdot{\bf D}-{\bf J}\cdot\nabla H \,{\bf N} = \nabla \cdot ({\bf J} H {\bf N})$ shares this structure.

\subsubsection{Hydrodynamic stresses \& forces}

The membrane is given to flow with a velocity ${\bf V} = {\bf v} + \mathrm{v}_\mathrm{n} {\bf N}$. The continuity equation for flows on the membrane is given as
\begin{equation}
    \nabla\cdot{\bf V} = \nabla \cdot {\bf v} + \mathrm{v}_{\mathrm{n}} \nabla \cdot {\bf N} = \nabla \cdot{\bf v} - 2H \mathrm{v}_\mathrm{n} = 0\text{,}
\end{equation}
which gives the classic result for the continuity equation on a deformable curved surface~\citep{arroyo2009}. The viscous stress tensor of a two-dimensional fluid membrane in the Newtonian limit is given by the purely tangential part of the symmetric velocity gradient 
\begin{equation}
    {\boldsymbol \sigma}^\mathrm{v} = \eta \bigl( {\bf I}_3 - {\bf NN} \bigr) \cdot \Bigl( \nabla {\bf V} + \bigl (\nabla {\bf V} \bigr)^{T} \Bigr) \cdot \bigl( {\bf I}_3 - {\bf NN} \bigr) = \eta \Bigl( \nabla_{\parallel} {\bf v} + \bigl( \nabla_{\parallel} {\bf v} \bigr)^T - 2 \mathrm{v}_{\mathrm{n}} {\bf S} \Bigr) \text{,}
    \label{SI:eq:hydroStress}  
\end{equation}
where $\eta$ is the membrane viscosity and $\nabla_{\parallel} {\bf v} = \nabla {\bf v} - (\nabla {\bf v} \cdot {\bf N}) {\bf N}$ is the covariant derivative of ${\bf v}$. The hydrodynamic force is given by the divergence of this stress 
\begin{equation}
    {\bf f}^{\textrm{v}} = \nabla \cdot \boldsymbol{\sigma}^{\textrm{v}} = \nabla_{\parallel} \cdot \boldsymbol{\sigma}^{\textrm{v}} + \Bigl[ \bigl( \nabla \cdot \boldsymbol{\sigma}^{\textrm{v}} \bigr) \cdot {\bf N} \Bigr] {\bf N} \text{,}
\end{equation}
and has both tangential and normal components. To simplify the normal component we use 
\begin{align}
    \bigl( \nabla \cdot \boldsymbol{\sigma}^{\textrm{v}} \bigr) \cdot {\bf N} & = \nabla \cdot \bigl( \boldsymbol{\sigma}^{\textrm{v}} \cdot {\bf N} \bigr) - \boldsymbol{\sigma}^{\textrm{v}} : \nabla {\bf N} , \\
    & = 2H \nabla \cdot {\bf v} + 2 \nabla_{\parallel} {\bf v} : {\bf D} - 2 \mathrm{v}_{\textrm{n}} \,{\bf S} : {\bf S} , \\
    & = 2H \nabla \cdot {\bf v} + 2 \nabla_{\parallel} {\bf v} : {\bf D} - 4 \mathrm{v}_{\textrm{n}} \bigl( 2H^2 - K_{\textrm{G}} \bigr) . 
\end{align}
In simplifying the tangential component of the force we make use of the standard result 
\begin{equation}
    \nabla_{\parallel} \cdot \Bigl( \bigl( \nabla_{\parallel} {\bf v} \bigr)^T \Bigr) = \nabla_{\parallel} \bigl( \nabla_{\parallel} \cdot {\bf v} \bigr) + K_{\textrm{G}} \,{\bf v} \text{,}
\end{equation}
obtained by commuting the order of covariant derivatives. Finally, using the continuity equation to replace $\nabla \cdot {\bf v}$ with $2H \,\mathrm{v}_{\textrm{n}}$ we find 
\begin{equation}
    {\bf f}^\mathrm{v} = \eta \Bigl( \nabla_{\parallel}^2 {\bf v} + K_{\textrm{G}} \,{\bf v} - 2 \mathrm{v}_{\textrm{n}} \nabla H - 2 {\bf D} \cdot \nabla \mathrm{v}_{\textrm{n}} \Bigr) + 2\eta \bigl[ \nabla_{\parallel} {\bf v} : {\bf D} - 2 \mathrm{v}_{\textrm{n}} \bigl( H^2 - K_{\textrm{G}} \bigr) \bigr] {\bf N} \text{.}
\end{equation}

\subsection{III. Calculational details for examples described in the main text}

\subsubsection{Odd mechanics of a spherical vesicle}

For a spherical membrane of radius $R$ the unperturbed geometry is isotropic; the shape operator is ${\bf S} = - \frac{1}{R} \,{\bf I}_2$ and the surface is totally umbilic with ${\bf D} = 0$. As a result, to linear order there is no Scriven-Love force and the spherical membrane is linearly stable. We focus on the linear description of the purely tangential flows ($\mathrm{v}_n = 0$) induced by a normal perturbation of the shape by an amount $\psi$. According to the general first variation formulae the mean curvature of the perturbed surface is 
\begin{equation}
    H = - \frac{1}{R} + \frac{1}{2} \nabla^2 \psi + \frac{1}{R^2} \psi + O(\psi^2) .
\end{equation} 
For purely tangential membrane flows, incompressibility gives $\nabla \cdot {\bf v} = 0$, whose solution is 
\begin{equation}
    {\bf v} = {\bf J} \cdot \nabla f ,
\end{equation}
for some function $f$. From a direct calculation (in a local basis) we find 
\begin{equation}
    \nabla_{\parallel}^2 {\bf v} = {\bf J} \cdot \nabla \bigl( \nabla^2 f \bigr) + \frac{1}{R^2} \,{\bf J} \cdot \nabla f .
\end{equation}
We can now write the tangential force balance as 
\begin{equation}
\eta \,{\bf J} \cdot \nabla \biggl( \nabla^2 f + \frac{2}{R^2} f \biggr) + \nabla \gamma + \zeta \,{\bf J} \cdot \nabla \biggl( \frac{1}{2} \nabla^2 \psi + \frac{1}{R^2} \psi \biggr) = 0 ,
\end{equation}
and it follows that $\gamma$ is constant and also that $(\nabla^2 + \frac{2}{R^2})(f + \frac{\zeta}{2\eta} \psi)$ is constant. It is natural to analyse the perturbations in terms of spherical harmonics, $Y_{l,m}$. For $l=0$ the perturbation $\psi$ is constant and there is no flow ($f=0$). For $l=1$ we find $(\nabla^2 + 2/R^2) \psi = 0$ and again there is no flow; this is a reflection of the fact that the $l=1$ modes are translations to linear order. For all higher multipoles the flow is non-trivial and given by 
\begin{equation}
f = - \frac{\zeta}{2\eta} \psi .
\end{equation}

To describe the response in a coordinate-free way, we adopt Maxwell's definition of the spherical harmonics in terms of derivatives of $\frac{1}{r}$~\cite{Maxwell,dennis2004,arnold1996}. Taking ${\bf a}_{1}, \dots , {\bf a}_{l}$ to be $l$ (unit) vectors in $\mathbb{R}^3$, an order $l$ multipole is given by  
\begin{equation}
\psi = \psi_l \biggl. R^{l+2} \nabla_{{\bf a}_1} \cdots \nabla_{{\bf a}_l} \frac{1}{r} \biggr|_{r=R} ,
\end{equation}
where $\psi_l$ is a dimensionless constant setting the amplitude. For instance, for the quadrupoles, $l=2$, we have 
\begin{equation}
\psi = \psi_2 R \biggl[ \frac{3 ({\bf a}_1 \cdot {\bf x}) ({\bf a}_2 \cdot {\bf x})}{R^2} - {\bf a}_{1} \cdot {\bf a}_2 \biggr]_{|{\bf x}| = R} ,
\end{equation}
and for the octupoles, $l=3$, we find  
\begin{equation}
\psi = \psi_3 \biggl[ - \frac{15 ({\bf a}_1 \cdot {\bf x}) ({\bf a}_2 \cdot {\bf x}) ({\bf a}_3 \cdot {\bf x})}{R^2} + 3 \bigl( {\bf a}_{1} \cdot {\bf a}_2 \bigr) \bigl( {\bf a}_3 \cdot {\bf x} \bigr) + 3 \bigl( {\bf a}_{2} \cdot {\bf a}_3 \bigr) \bigl( {\bf a}_1 \cdot {\bf x} \bigr) + 3 \bigl( {\bf a}_{3} \cdot {\bf a}_1 \bigr) \bigl( {\bf a}_2 \cdot {\bf x} \bigr) \biggr]_{|{\bf x}| = R} .
\end{equation}
The associated flows are ${\bf v} = - \frac{\zeta}{2\eta} \,{\bf J} \cdot \nabla \psi$. We make use of the relation ${\bf X} = R \,{\bf N}$ for the sphere and write these as 
\begin{align}
{\bf v} & = - \frac{3\zeta}{2\eta} \,\psi_2 \Bigl[ \bigl( {\bf a}_1 \cdot {\bf N} \bigr) \,{\bf N} \times {\bf a}_2 + \bigl( {\bf a}_2 \cdot {\bf N} \bigr) \,{\bf N} \times {\bf a}_1 \Bigr] , \\
\begin{split}
{\bf v} & = \frac{3\zeta}{2\eta} \,\psi_3 \biggl[ \Bigl( 5 \bigl( {\bf a}_1 \cdot {\bf N} \bigr) \bigl( {\bf a}_2 \cdot {\bf N} \bigr) - {\bf a}_1 \cdot {\bf a}_2 \Bigr) {\bf N} \times {\bf a}_3 + \Bigl( 5 \bigl( {\bf a}_2 \cdot {\bf N} \bigr) \bigl( {\bf a}_3 \cdot {\bf N} \bigr) - {\bf a}_2 \cdot {\bf a}_3 \Bigr) {\bf N} \times {\bf a}_1 \\
& \quad \qquad + \Bigl( 5 \bigl( {\bf a}_3 \cdot {\bf N} \bigr) \bigl( {\bf a}_1 \cdot {\bf N} \bigr) - {\bf a}_3 \cdot {\bf a}_1 \Bigr) {\bf N} \times {\bf a}_2 \biggr] ,
\end{split}
\end{align}
for the quadrupoles ($l=2$) and octupoles ($l=3$), respectively. We used these general results to generate Fig.~\ref{fig:spheroidFlows}(a) of the main text. 

\subsubsection{Spheroidal Membrane}

In this section we summarise the calculational details for the tangential flows on a spheroidal membrane with arbitrary aspect ratio. We take the membrane to have the shape of an ellipsoid with principal semi-axes $a,a,c$ and embedding 
\begin{equation}
    {\bf X} = \bigl( a \sin\theta \cos\phi , a \sin\theta \sin\phi , c \cos\theta \bigr) ,
\end{equation}
where $\theta,\phi$ are standard spherical polar angles. We find 
\begin{equation}
    d{\bf X} = {\bf e}_{\theta} \,\bigl( a^2 + (c^2-a^2) \sin^2\theta \bigr)^{1/2} \,d\theta + {\bf e}_{\phi} \,a \sin\theta \,d\phi ,
\end{equation}
where 
\begin{align}
    {\bf e}_{\theta} & = \Bigl( a^2 + (c^2-a^2) \sin^2\theta \Bigr)^{-1/2} \,\bigl[ a \cos\theta \cos\phi , a \cos\theta \sin\phi , -c \sin\theta \bigr] , \\
    {\bf e}_{\phi} & = \bigl[ - \sin\phi , \cos\phi , 0 \bigr] ,
\end{align}
are an orthonormal basis for the tangent space. The unit outward normal is 
\begin{equation}
    {\bf N} = \Bigl( a^2 + (c^2-a^2) \sin^2\theta \Bigr)^{-1/2} \,\bigl[ c \sin\theta \cos\phi , c \sin\theta \sin\phi , a \cos\theta \bigr] ,
\end{equation}
and the shape operator is 
\begin{equation}
    {\bf S} = - \nabla {\bf N} = \Bigl( a^2 + (c^2-a^2) \sin^2\theta \Bigr)^{-3/2} \frac{c}{a} \biggl\{ - \biggl( a^2 + \frac{1}{2} (c^2-a^2) \sin^2\theta \biggr) {\bf I}_2 + \frac{1}{2} (c^2-a^2) \sin^2\theta \bigl[ {\bf e}_{\theta} {\bf e}_{\theta} - {\bf e}_{\phi} {\bf e}_{\phi} \bigr] \biggr\} .
\end{equation}
We can read off the mean curvature 
and then a short direct calculation gives 
\begin{equation}
    {\bf J} \cdot \nabla H = \Bigl( a^2 + (c^2-a^2) \sin^2\theta \Bigr)^{-3} \biggl( a^2 + \frac{1}{4} (c^2-a^2) \sin^2\theta \biggr) \frac{c}{a} (c^2-a^2) \sin 2\theta \,{\bf e}_{\phi} .
\end{equation}

For the membrane flow, we look for a solution with purely azimuthal velocity ${\bf v} = v(\theta) \,{\bf e}_{\phi}$. The gradient is 
\begin{equation}
    \nabla {\bf v} = \bigl( a^2 + (c^2-a^2) \sin^2\theta \bigr)^{-1/2} \biggl( \partial_{\theta} v \,{\bf e}_{\theta} {\bf e}_{\phi} - \frac{\cos\theta}{\sin\theta} \,v \,{\bf e}_{\phi} {\bf e}_{\theta} - \frac{c}{a} \,v \,{\bf e}_{\phi} {\bf N} \biggr) ,
\end{equation}
and we then calculate directly that 
\begin{equation}
    \begin{split}
        \nabla_{\parallel} \cdot \Bigl( \nabla_{\parallel} {\bf v} + ( \nabla_{\parallel} {\bf v} )^T \Bigr) & = \bigl( a^2 + (c^2-a^2) \sin^2\theta \bigr)^{-1} \biggl( \frac{1}{\sin\theta} \partial_{\theta} \Bigl( \sin\theta \,\partial_{\theta} v \Bigr) - \frac{1}{\sin^2\theta} v + 2v \biggr) {\bf e}_{\phi} \\
        & \quad - \bigl( a^2 + (c^2-a^2) \sin^2\theta \bigr)^{-2} (c^2-a^2) \sin\theta \cos\theta \biggl( \partial_{\theta} v - \frac{\cos\theta}{\sin\theta} v \biggr) {\bf e}_{\phi} .
    \end{split}
\end{equation}
It is clear that $\nabla {\bf v} : {\bf D} = 0$ so that there is no Scriven-Love term and we only have to solve the tangential force balance, which reduces to 
\begin{multline}
    \frac{1}{\sin\theta} \partial_{\theta} \Bigl( \sin\theta \,\partial_{\theta} v \Bigr) - \frac{1}{\sin^2\theta} v + 2v - \frac{(c^2-a^2) \sin\theta \cos\theta}{a^2 + (c^2-a^2) \sin^2\theta} \biggl( \partial_{\theta} v - \frac{\cos\theta}{\sin\theta} v \biggr) \\
    = - \frac{2\zeta}{\eta} \Bigl( a^2 + (c^2-a^2) \sin^2\theta \Bigr)^{-2} \biggl( a^2 + \frac{1}{4} (c^2-a^2) \sin^2\theta \biggr) \frac{c}{a} (c^2-a^2) \sin\theta \cos\theta .
\end{multline}
The solution for the boundary conditions $v(0)=v(\pi)=0$ is given by 
\begin{equation}
    v(\theta) = \begin{cases}
    \frac{\zeta}{2\eta} \Bigl( \frac{c^2}{a^2} - 1 \Bigr)^{1/2} \,\textrm{arctanh} \biggl[ \Bigl( 1 - \frac{a^2}{c^2} \Bigr)^{1/2} \cos\theta \biggr] \sin\theta , & \frac{c}{a} > 1 \quad (\textrm{prolate}) , \\[3mm]
    - \frac{\zeta}{2\eta} \Bigl( 1 - \frac{c^2}{a^2} \Bigr)^{1/2} \,\arctan \biggl[ \Bigl( \frac{a^2}{c^2} - 1 \Bigr)^{1/2} \cos\theta \biggr] \sin\theta , & \frac{c}{a} < 1 \quad (\textrm{oblate}) ,
    \end{cases}
\end{equation}
and is plotted in Fig.~\ref{fig:spheroidFlows}(b). Of particular note is the change in sign, and hence directionality, of the flow between the prolate and oblate cases. Finally, linearising in $(c/a)-1$ the solution (for both prolate and oblate spheroids) reduces to  
\begin{equation}
    v(\theta) = \frac{\zeta}{\eta} \biggl( \frac{c}{a} - 1 \biggr) \sin\theta \cos\theta .
\end{equation}

\subsubsection{Instability of an odd active membrane tube}

In this section we summarise the calculation of the odd mechanical shape instability for a membrane tube. We take the unperturbed membrane to be the cylinder ${\bf X}(\theta,z) = (r\cos\theta , r\sin\theta , z)$ of radius $r$, for which the shape operator is ${\bf S} = - \frac{1}{r} \,{\bf e}_{\theta} {\bf e}_{\theta}$. Perturbing the surface by an amount $\psi(\theta,z,t)$ along its normal, the general first variation formulae give the mean and Gaussian curvature as 
\begin{align}
    & H = -\frac{1}{2r} + \frac{1}{2} \biggl( \partial_{zz} \psi + \frac{1}{r^2} \partial_{\theta\theta} \psi + \frac{1}{r^2} \psi \biggr) \text{,} 
    && K_{\text{G}} = -\frac{1}{r} \,\partial_{zz} \psi \text{.}
\end{align}
We then compute the odd active force density to be 
\begin{equation}
    \zeta \,{\bf J} \cdot \nabla H = \frac{\zeta}{2} \biggl[ {\bf e}_2 \,\frac{1}{r} \partial_{\theta} - {\bf e}_1 \,\partial_{z} \biggr] \biggl( \partial_{zz} \psi + \frac{1}{r^2} \partial_{\theta\theta} \psi + \frac{1}{r^2} \psi \biggr) .
\end{equation}

\begin{figure}
    \centering
    \includegraphics[width=0.8\textwidth]{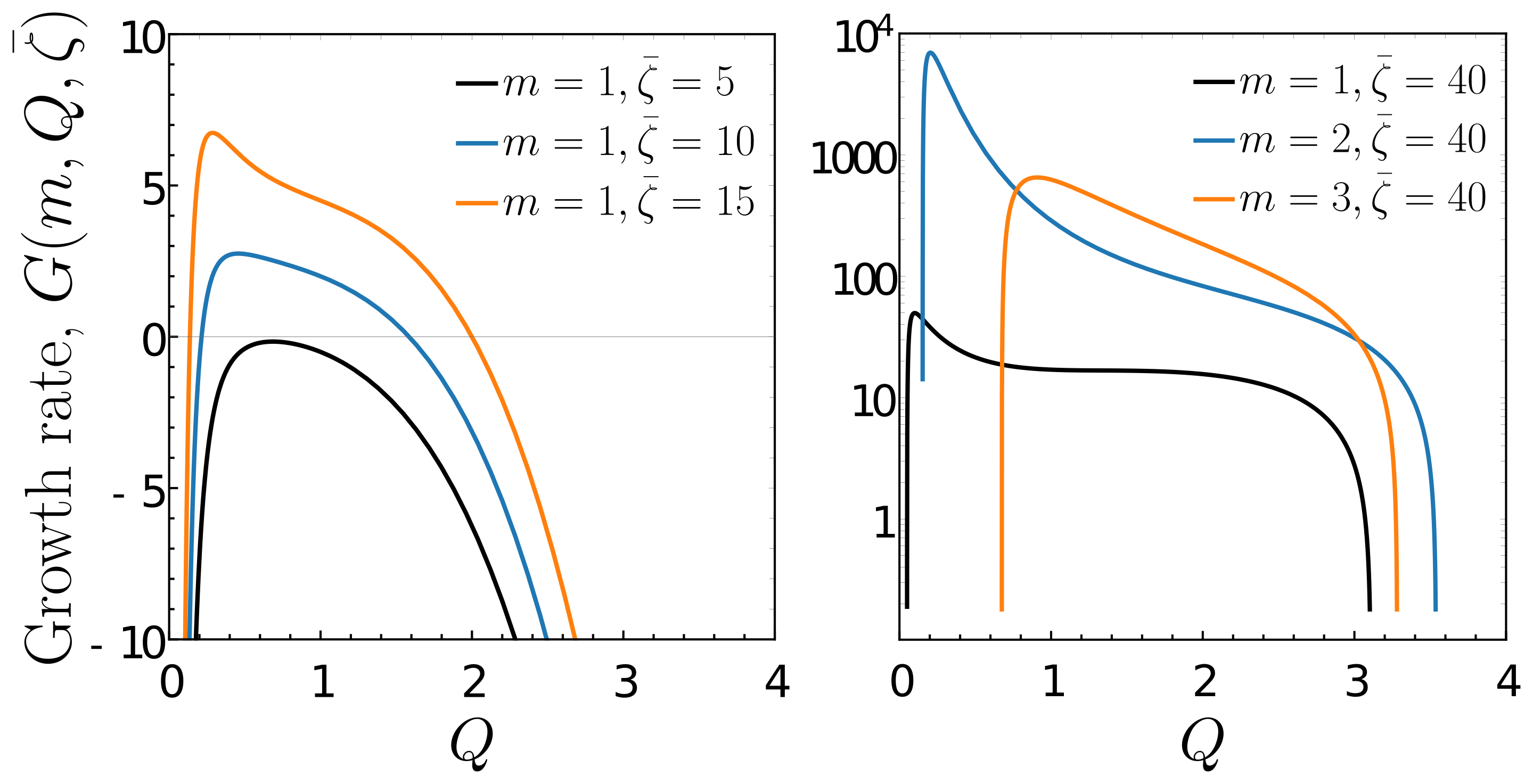}
    \caption{Growth rate for a perturbed tube. Left shows the growth for the left handed helical ($m=1$) mode for varying active Scriven-Love number $\bar\zeta$. Right shows the growth for a fixed $\bar \zeta$ but for different values of $m$, \textit{i.e.}~single, double and triple helical deformations. As active stress is increased modes of higher $m$ value become more unstable.}
    \label{SI:fig:tubeGrowthRate}
\end{figure}

We write the membrane velocity as ${\bf V} = \mathrm{v}_1 {\bf e}_1 + \mathrm{v}_2 {\bf e}_2 + \mathrm{v}_\textrm{n} {\bf N}$. To linear order, incompressibility gives 
\begin{equation}
    \frac{1}{r} \,\partial_{\theta} \mathrm{v}_1 + \partial_z \mathrm{v}_2 + \frac{\mathrm{v}_{\textrm{n}}}{r} = 0 , \qquad \Rightarrow \quad \mathrm{v}_{\textrm{n}} = - \partial_{\theta} \mathrm{v}_1 - r \,\partial_z \mathrm{v}_2 ,
    \label{SI:eq:tube_continuity}
\end{equation}
and we find the hydrodynamic force is 
\begin{equation}
    \begin{split}
        {\bf f}^{\text{v}} & = \frac{\eta \left( r \left(\partial_{z\theta} \mathrm{v}_2 + r \partial_{zz} \mathrm{v}_1 \right)+2 \left(\partial_{\theta\theta}\mathrm{v}_1 + \partial_{\theta} \mathrm{v}_\textrm{n}\right)\right)}{r^2} {\bf e}_1 + \frac{\eta    \left(r \left(\partial_{z\theta} \mathrm{v}_1 + 2 r \partial_{zz} \mathrm{v}_2\right)+ \partial_{\theta\theta} \mathrm{v}_2\right)}{r^2} {\bf e}_2 \\
        & \quad - \frac{2 \eta \left( \partial_{\theta} \mathrm{v}_1 + \mathrm{v}_\textrm{n}\right)}{r^2} {\bf N} \text{.}
    \end{split}
\end{equation}
Balancing the tangential components against the chiral force gives 
\begin{align}
    &{\bf e}_1: \frac{\eta  \left( r \,\partial_{z\theta} \mathrm{v}_2 + r^2 \partial_{zz} \mathrm{v}_1 + 2 \partial_{\theta\theta} \mathrm{v}_1 + 2 \partial_\theta \mathrm{v}_\textrm{n} \right)}{r^2} - \frac{\zeta    \left(\partial_{z\theta\theta} \psi + r^2 \partial_{zzz} \psi + \partial_z \psi \right)}{2 r^2} + \frac{1}{r} \partial_\theta \gamma = 0\text{,}\\
    &{\bf e}_2: \frac{\eta  \left( r \,\partial_{z\theta} \mathrm{v}_1 + 2 r^2 \partial_{zz} \mathrm{v}_2 + \partial_{\theta\theta} \mathrm{v}_2 \right)}{r^2} + \frac{\zeta  \left(r^2 \partial_{zz\theta} \psi + \partial_{\theta\theta\theta} \psi + \partial_\theta \psi\right)}{2 r^3} +  \,\partial_z \gamma = 0 \text{,}
\end{align}
while the normal force balance yields the shape equation 
\begin{equation}
    \biggl( \gamma - \frac{\kappa}{2r^2} \biggr) \biggl( - \frac{1}{r} + \frac{1}{r^2} \bigl( 1 + \partial_{\theta\theta} + r^2 \partial_{zz} \bigr) \psi \biggr) - \frac{ \kappa}{r^4} \Bigl[ \bigl( 1 + \partial_{\theta\theta} + r^2 \partial_{zz} \bigr)^2 - 2r^2 \partial_{zz} \Bigr] \psi - \frac{2\eta}{r^2} \bigl( \partial_{\theta} \mathrm{v}_1 + \mathrm{v}_{\textrm{n}} \bigr) = 0\text{,}
    \label{SI:eq:tube_shape}
\end{equation} 
The unperturbed membrane tube is in mechanical equilibrium if the surface tension is $\gamma = \kappa/(2r^2)$; for the perturbed surface $\gamma$ will deviate from this value by an amount linear in $\psi$. 
Eliminating the active terms between the two tangential balances we obtain an equation for the surface tension 
\begin{equation}
    \biggl( \frac{1}{r^2} \partial_{\theta\theta} + \partial_{zz} \biggr) \gamma = - 2\eta \biggl( \frac{1}{r^2} \,\partial_{\theta\theta} + \partial_{zz} \biggr) \biggl( \frac{1}{r} \,\partial_{\theta} \mathrm{v}_1 + \partial_z \mathrm{v}_2 \biggr) - \frac{2\eta}{r^3} \,\partial_{\theta\theta} \mathrm{v}_{\textrm{n}} ,
    \label{SI:eq:tube_tension}
\end{equation}
while by eliminating the surface tension we obtain 
\begin{equation}
    \eta \biggl( \frac{1}{r^2} \,\partial_{\theta\theta} + \partial_{zz} \biggr) \biggl( \frac{1}{r} \,\partial_{\theta} \mathrm{v}_2 - \partial_{z} \mathrm{v}_1 \biggr) - \frac{2\eta}{r^2} \,\partial_{\theta z} \mathrm{v}_{\textrm{n}} = - \frac{\zeta}{2} \biggl( \frac{1}{r^2} \,\partial_{\theta\theta} + \partial_{zz} \biggr) \biggl( \frac{1}{r^2} \bigl( \psi + \partial_{\theta\theta} \psi \bigr) + \,\partial_{zz} \psi \biggr) .
    \label{SI:eq:tube_flow}
\end{equation}
After replacing $\mathrm{v}_{\textrm{n}}$ with the tangential components of flow using the continuity equation~\eqref{SI:eq:tube_continuity}, the equations~\eqref{SI:eq:tube_tension},~\eqref{SI:eq:tube_flow} and~\eqref{SI:eq:tube_shape} may be considered to give a system that solves for the surface tension and tangential components of flow in terms of the normal surface displacement $\psi$. Finally, the membrane dynamics is obtained from incompressibility to give $\mathrm{v}_{\textrm{n}}$ in terms of $\psi$ and the kinematic condition $\mathrm{v}_{\textrm{n}} = \partial_t \psi$. The analysis is facilitated by Fourier transforming $(\theta,z) \to (m,q)$ and gives the mode evolution equation 
\begin{equation}
    \partial_t \tilde{\psi}_{m}(q) = \biggl\{ \frac{\zeta}{\eta r} \frac{m qr (m^2+q^2r^2) (m^2-1+q^2r^2)}{4 q^4r^4} - \frac{\kappa}{\eta r^2} \frac{(m^2+q^2r^2)^2}{4q^4r^4} \Bigl[ \bigl( m^2 - 1 + q^2 r^2 \bigr)^2 + 2 q^2 r^2 \Bigr] \biggr\} \tilde{\psi}_{m}(q) .
\end{equation}
We pass to dimensionless variables by setting $Q = qr$, $\bar{\zeta} = \zeta r/\kappa$ and $\bar{t} = (\eta r^2/\kappa) t$ to obtain 
\begin{equation}
    \partial_{\bar{t}} \tilde{\psi}_{m} = \frac{m^2+Q^2}{4Q^4} \Bigl[ \bar{\zeta} mQ \bigl( m^2 - 1 + Q^2 \bigr) - \bigl( m^2 + Q^2 \bigr) \bigl( m^4 + 2m^2 (Q^2 - 1) + Q^4 + 1 \bigr) \Bigr] \tilde{\psi}_{m} ,
\end{equation}
with growth rate $G_m(Q)$ given in Eq.~(11) 
of the main text. This growth rate is plotted for several values of $\bar{\zeta}$ in \fig{SI:fig:tubeGrowthRate}. 

\subsubsection{Instability of a helicoid}
\label{SI:subsec:helicoid}

We parameterise the helicoid as 
\begin{equation}
    {\bf X}(u,v) = \biggl( \frac{\sinh \alpha v}{\alpha} \cos \alpha u , \frac{\sinh \alpha v}{\alpha} \sin \alpha u , u \biggr) ,
\end{equation}
where $\alpha$ is a wavevector. The metric is $ds^2 = \cosh^2 \alpha v \bigl( du^2 + dv^2 \bigr)$ and a right-handed orthonormal frame adapted to the surface is 
\begin{align}
    {\bf e}_u & = \bigl[ - \tanh \alpha v \sin \alpha u , \tanh \alpha v \cos \alpha u , \textrm{ sech}\, \alpha v \bigr] , \\
    {\bf e}_v & = \bigl[ \cos \alpha u , \sin \alpha u , 0 ] , \\
    {\bf N} & = \bigl[ - \textrm{ sech}\, \alpha v \sin \alpha u , \textrm{sech}\, \alpha v \cos \alpha u , - \tanh \alpha v \bigr] .
\end{align}
The shape operator is 
\begin{equation}
{\bf S} = - \nabla {\bf N} = \alpha \,\textrm{ sech}^2 \alpha v \bigl( {\bf e}_u {\bf e}_v + {\bf e}_v {\bf e}_u \bigr) ,
\end{equation}
and the principal curvature directions are ${\bf E}_1 = ({\bf e}_u + {\bf e}_v)/\sqrt{2}$ and ${\bf E}_2 = ({\bf e}_v - {\bf e}_u)/\sqrt{2}$. The mean curvature vanishes (as the helicoid is a minimal surface) and the Gaussian curvature is 
\begin{equation}
K_{\textrm{G}} = - \alpha^2 \,\mathrm{sech}^4 \alpha v .
\end{equation}

We now deform the surface by displacing it by an amount $\psi(u,v,t)$ along its normal and use the general first normal variation formulae to compute the new geometry.  
The variation in mean curvature is given by 
\begin{equation}
H = \frac{1}{2} \nabla^2 \psi + \alpha^2 \,\mathrm{sech}^4 \alpha v \,\psi + O(\psi^2) = \frac{1}{2} \bigl( \nabla^2 - 2 K_{\textrm{G}} \bigr) \psi + O(\psi^2) ,
\end{equation}
and it follows that, to first order in $\psi$, 
\begin{equation}
\nabla^2 H + 2H \bigl( H^2 - K_{\textrm{G}} \bigr) = \frac{1}{2} \bigl( \nabla^2 - 2 K_{\textrm{G}} \bigr)^2 \psi + O(\psi^2) .
\end{equation}

For the membrane flow we write as usual ${\bf V} = {\bf v} + \mathrm{v}_{\textrm{n}} \,{\bf N}$ and have the general form of the continuity equation $\nabla \cdot {\bf v} = 2H \mathrm{v}_{\textrm{n}}$, but if we work to first order then this reduces to $\nabla \cdot {\bf v} = 0$, {\sl i.e.} the tangential part of the membrane flow is incompressible. The general solution is 
\begin{equation}
{\bf v} = {\bf J} \cdot \nabla f ,
\end{equation}
for some function $f$. From a direct calculation we find 
\begin{equation}
    \nabla_{\parallel}^2 \bigl( {\bf J}\cdot\nabla f \bigr) = {\bf J} \cdot \nabla \bigl( \nabla^2 f \bigr) - \alpha^2 \,\mathrm{sech}^4 \alpha v \,{\bf J} \cdot \nabla f\mathrm{,}
\end{equation}
and can then write the tangential and normal force balances 
\begin{gather}
    \begin{split}
    & \eta \Bigl[ {\bf J} \cdot \nabla \bigl( \nabla^2 f \bigr) - 2 \alpha^2 \,\mathrm{sech}^4 \alpha v \,{\bf J} \cdot \nabla f - 2 \alpha \,\textrm{sech}^3 \alpha v \bigl( {\bf e}_u \,\partial_v \mathrm{v}_{\textrm{n}} + {\bf e}_v \,\partial_u \mathrm{v}_{\textrm{n}} \bigr) \Bigr] \\
    & \quad + \nabla \gamma + \frac{\zeta}{2} \,{\bf J} \cdot \nabla \Bigl( \bigl( \nabla^2 + 2 \alpha^2 \,\mathrm{sech}^4 \alpha v \bigr) \psi \Bigr) = 0 ,
    \end{split} \\
    \begin{split}
    & \gamma \bigl( \nabla^2 + 2 \alpha^2 \,\mathrm{sech}^4 \alpha v \bigr) \psi - \kappa \bigl( \nabla^2 + 2 \alpha^2 \,\mathrm{sech}^4 \alpha v \bigr)^2 \psi \\
    & + 2\eta \Bigl[ \alpha \,\textrm{sech}^2 \alpha v \bigl( \partial_{uu} f - \partial_{vv} f + 2 \alpha \tanh \alpha v \,\partial_v f \bigr) - 2 \alpha^2 \,\mathrm{sech}^4 \alpha v \,\mathrm{v}_{\textrm{n}} \Bigr] = 0 .
    \end{split}
\end{gather}

From here we develop only the case where there is invariance along the axis of the helicoid and there is no dependence on $u$. Then the ${\bf e}_v$ component of the tangential force balance gives that the tension $\gamma$ is constant, to linear order, while the ${\bf e}_u$ component gives 
\begin{equation}
\begin{split}
& \partial_v \bigl( \mathrm{sech}^2 \alpha v \,\partial_{vv} f \bigr) - 2 \alpha^2 \,\mathrm{sech}^4 \alpha v \,\partial_v f + 2 \alpha \,\textrm{sech}^2 \alpha v \,\partial_v v_n + \frac{\zeta}{2\eta} \,\partial_v \Bigl( \mathrm{sech}^2 \alpha v \bigl( \partial_{vv} + 2 \alpha^2 \,\mathrm{sech}^2 \alpha v \bigr) \psi \Bigr) = 0 .
\end{split}
\end{equation}
From the normal force balance we obtain the normal component of velocity as 
\begin{equation}
    \begin{split}
    \mathrm{v}_{\textrm{n}} & = \frac{\gamma}{4\eta \alpha^2} \cosh^2 \alpha v \bigl( \partial_{vv} + 2 \alpha^2 \,\mathrm{sech}^2 \alpha v \bigr) \psi - \frac{\kappa}{4\eta \alpha^2} \cosh^4 \alpha v \bigl( \mathrm{sech}^2 \alpha v \,\partial_{vv} + 2 \alpha^2 \,\mathrm{sech}^4 \alpha v \bigr)^2 \psi \\
    & \quad - \frac{1}{2\alpha} \cosh^4 \alpha v \,\partial_v \bigl( \mathrm{sech}^2 \alpha v \,\partial_{v} f \bigr) \text{.}
    \end{split}
\label{SI:eq:helicoidNormalVelocity}
\end{equation}

In order to make progress here we can consider a simple numerical bootstrapping procedure. To do this we neglect the contribution to the tangential force balance from the normal velocity gradients and solve the following equation numerically
\begin{equation}
\partial_v \bigl( \mathrm{sech}^2 \alpha v \,\partial_{vv} f \bigr) - 2 \alpha^2 \,\mathrm{sech}^4 \alpha v \,\partial_v f + \frac{\zeta}{2\eta} \,\partial_v \Bigl( \mathrm{sech}^2 \alpha v \bigl( \partial_{vv} + 2 \alpha^2 \,\mathrm{sech}^2 \alpha v \bigr) \psi \Bigr) = 0 \text{,}
\end{equation}
for a deformation defined by $\psi = \delta\cos(\pi v/2L)$, where we choose $L=\alpha=1$. This is solved for $f$ using NDSolve in Mathematica with boundary conditions $f(L)=0$, $f'(L)=f'(-L)=0$ to give no flow at the frame of the helicoid. The solution is then substituted directly into~\eqref{SI:eq:helicoidNormalVelocity} for the normal velocity to produce the figures in the main text.

In order to compare to the case of a soap film we can make a simplifed approximation to the equations. Neglecting tangential flow and Fourier transforming $v\to q$ and expanding around $v=0$ we can reduce~\eqref{SI:eq:helicoidNormalVelocity} to
\begin{equation}
    \partial_t \bar{\psi}_q \approx \frac{\gamma}{4\eta \alpha^2}  \left(2 \alpha^2-q^2\right)  \bar{\psi}_q -  \frac{\kappa}{4\eta \alpha^2} \left(2 \alpha^2-q^2\right)^2\bar{\psi}_q \text{.}
\end{equation}
In the absence of activity this captures the well-known instability of the helicoid~\cite{boudaoud1999,machon2016PRL,alexander2020}, albeit for a critical size $q_c = \sqrt{2}\alpha \Rightarrow \alpha v_c = \pi/(2\sqrt{2}) \approx 1.11$ that is a little smaller than the exact value, $\approx 1.19968$. The bending rigidity also introduces a finite size to the linear growth rate of $q^* = \sqrt{2\alpha^2-\gamma/2\kappa}$.

Numerically we find that the activity produces instability of the otherwise stable helicoid when $\zeta\alpha\lessapprox-6\gamma$. The exact value of the critical activity for instability will be a little different but the coupling of the sign of $\zeta$ and handedness of the helicoid will be robust.

\subsubsection{Multiple helicoids and ``parking garage'' structures}

A well-known extension of the helicoid are the surfaces composed of an array of helicoidal axes, or screw dislocations, given parametrically by ${\bf X} = \bigl( x, y, h(x+iy) \bigr)$ for a `height function' $h$. The helicoid is recovered by the choice $h = (1/\alpha) \arctan(y/x)$. By taking superpositions of helicoids at different $(x,y)$ positions we obtain `parking garage' structures, ramps and twist-grain boundaries. These surfaces have been used to model structures in the rough endoplasmic reticulum (ER)~\cite{teresaki2013,guven2014} and plant photosynthetic membrane~\cite{bussi2019}. 

We give the calculation explicitly for the simplest case of a pair of helicoidal axes with opposite handedness, creating ramps between the layers in stacked membrane sheets such as the ER. The geometry is conveniently described using bipolar coordinates $(\xi,\eta)$ for the $xy$-plane 
\begin{align}
    & \xi + i \eta = \ln \frac{x+iy-R}{x+iy+R} , && x + iy = \frac{-R\sinh\xi}{\cosh\xi - \cos\eta} + i \frac{R\sin\eta}{\cosh\xi - \cos\eta} .
\end{align}
The helical axes are located at $x=\pm R$, $y=0$. We will also make use of the orthonormal basis for the bipolar coordinate system 
\begin{align}
    & {\bf e}_{\xi} = \frac{\cosh\xi \cos\eta - 1}{\cosh\xi - \cos\eta} \,{\bf e}_x - \frac{\sinh\xi \sin\eta}{\cosh\xi - \cos\eta} \,{\bf e}_y , 
    && {\bf e}_{\eta} = \frac{\sinh\xi \sin\eta}{\cosh\xi - \cos\eta} \,{\bf e}_x + \frac{\cosh\xi \cos\eta - 1}{\cosh\xi - \cos\eta} \,{\bf e}_y .
\end{align}
Using the bipolar coordinate system the surface can be parameterised as 
\begin{equation}
    {\bf X}(\xi,\eta) = \Bigl( x(\xi,\eta) , y(\xi,\eta) , p_0 \eta \Bigr) ,
\end{equation}
where $p_0$ is a constant setting the helical pitch. Infinitesimal displacements on the surface are given by 
\begin{equation}
    d{\bf X} = \bigl[ {\bf e}_{\xi} , 0 \bigr] \,\frac{R}{\cosh\xi - \cos\eta} \,d\xi + \biggl[ {\bf e}_{\eta} , \frac{p_0 (\cosh\xi - \cos\eta)}{R} \biggr] \,\frac{R}{\cosh\xi - \cos\eta} \,d\eta ,
\end{equation}
which gives an orthonormal basis for the tangent space to the surface 
\begin{align}
    & {\bf e}_1 = \bigl[ {\bf e}_{\xi} , 0 \bigr] , 
    && {\bf e}_2 = \biggl( 1 + \frac{p_0^2 (\cosh\xi - \cos\eta)^2}{R^2} \biggr)^{-1/2} \biggl[ {\bf e}_{\eta} , \frac{p_0 (\cosh\xi - \cos\eta)}{R} \biggr] ,
\end{align}
and the unit normal 
\begin{equation}
    {\bf N} = \biggl( 1 + \frac{p_0^2 (\cosh\xi - \cos\eta)^2}{R^2} \biggr)^{-1/2} \biggl[ - \frac{p_0 (\cosh\xi - \cos\eta)}{R} {\bf e}_{\eta} , 1 \biggr] .
\end{equation}
The shape operator is given by 
\begin{equation}
    \begin{split}
        {\bf S} & = - \frac{(p_0 \Omega)^3 \sin\eta}{2R} \Bigl( 1 + \bigl( p_0 \Omega \bigr)^2 \Bigr)^{-3/2} \bigl[ {\bf e}_1 {\bf e}_1 + {\bf e}_2 {\bf e}_2 \bigr] - \frac{p_0 \Omega \sin\eta}{2R} \Bigl( 2 + \bigl( p_0 \Omega \bigr)^2 \Bigr) \Bigl( 1 + \bigl( p_0 \Omega \bigr)^2 \Bigr)^{-3/2} \bigl[ {\bf e}_1 {\bf e}_1 - {\bf e}_2 {\bf e}_2 \bigr] \\
        & \quad + \frac{p_0 \Omega \sinh\xi}{R} \Bigl( 1 + \bigl( p_0 \Omega \bigr)^2 \Bigr)^{-1} \bigl[ {\bf e}_1 {\bf e}_2 + {\bf e}_2 {\bf e}_1 \bigr] ,
    \end{split}
\end{equation}
where we write $\Omega = (\cosh\xi - \cos\eta)/R$ for brevity, and we then calculate the chiral active force density from  
\begin{equation}
    \begin{split}
        {\bf J} \cdot \nabla H = \frac{3}{2R^2} \biggl( \frac{p_0 \Omega}{1+(p_0 \Omega)^2} \biggr)^3 & \biggl\{ {\bf e}_1 \biggl[ \sin^2\eta + \frac{1+(p_0 \Omega)^2}{3} \bigl( \cosh\xi - \cos\eta \bigr) \cos\eta \biggr] \\
        & \quad - {\bf e}_2 \Bigl( 1 + \bigl( p_0 \Omega \bigr)^2 \Bigr)^{1/2} \sinh\xi \sin\eta \biggr\} .
    \end{split}
\end{equation}
We use this expression for the chiral force density to produce Fig.~\ref{fig:oddHelicoid}(c) of the main text.

\begin{figure}
    \centering
    \includegraphics[width=0.9\textwidth]{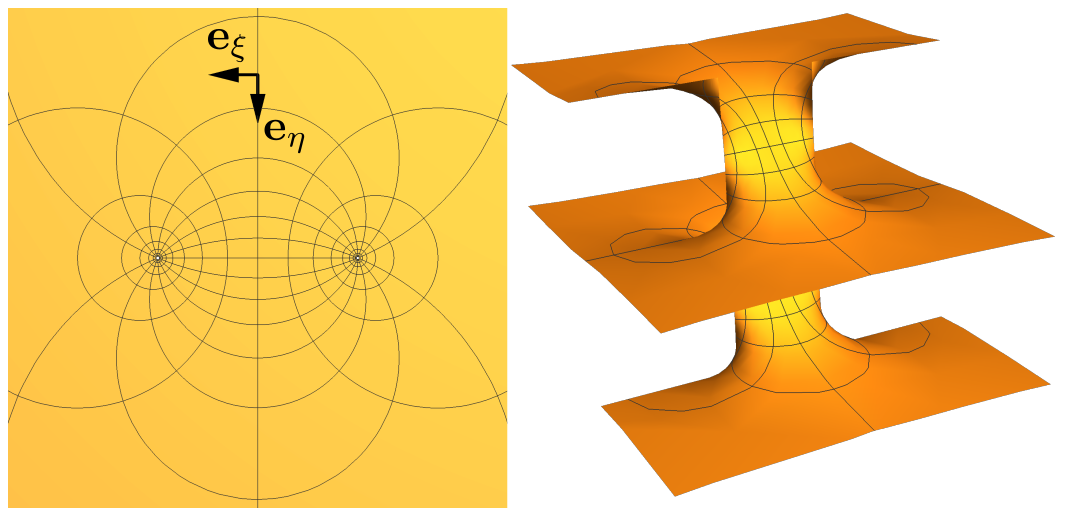}
    \caption{Helicoidal ``parking garage'' ramps in bipolar coordinates. \textbf{Left:} bipolar coordinates in $\mathbb{R}^2$ with the orthonormal basis $\{{\bf e}_\xi,{\bf e}_\eta\}$ labelled. \textbf{Right:} Helicoidal ramp of opposite handedness plotted for pitch $p_0=1$ and separations $R=2$. Contours on the surface denote the lines of constant bipolar coordinate.}
    \label{fig:ramps}
\end{figure}

%% file: main.bbl
\begin{thebibliography}{99}
\bibitem{note_LeviCivita} The complex structure is equivalent to the Levi-Civita tensor, differing only by a sign according to conventions.
\end{thebibliography}
